\documentclass[prb,superscriptaddress,twocolumn,showpacs,preprintnumbers,amsmath,amssymb,fixfloat]{revtex4-1}

\usepackage{graphicx}% Include figure files
\usepackage{dcolumn}% Align table columns on decimal point
\usepackage{bm}% bold math 0.011{\rm ~eV}
\usepackage{bbm}

\begin{document}
\title{Higher order moments, cumulants, and spectra of continuous quantum noise measurements}
\author{Daniel H\"agele}
\address{Ruhr University Bochum, Faculty of Physics and Astronomy, Experimental Physics VI (AG), Germany}
 \author{Fabian Schefczik}
 \address{Ruhr University Bochum, Faculty of Physics and Astronomy, Experimental Physics VI (AG), Germany}
\date{\today}
\begin{abstract}
We present general quantum mechanical expressions for higher order moments, cumulants, and spectra of continuously measured quantum systems with applications in spin noise spectroscopy, quantum transport, and measurement theory in general.
 Starting from the so-called
stochastic master equation of continuous measurement theory, we find that the leading orders of the fluctuating detector output $z(t)$ with respect to the measurement strength $\beta$ are a white shot noise background, a constant measurement offset, and the leading 
order quantum noise of the measured operator $A$. Starting from quantum expressions  for the multi-time moments
$\langle z(t_n)\cdots z(t_1) \rangle$ we derive three- and four-time cumulants that are valid in all orders of $\beta$ covering the full regime
between the weak and strong measurement limit (Zeno-limit). Intriguingly, quantum expressions for the cumulants were found that
exhibit the same simple structure as those for the moments after introduction of only a slightly modified system propagator.
Very compact expressions for the cumulant-based third and fourth order spectra (bispectrum and trispectrum) follow naturally. 
 We illustrate the usefulness of higher order spectra 
by treating a real world two-spin system with strong hyperfine interaction.
Moreover, spin noise spectroscopy is shown to have the potential for investigating the transition from weak measurements to the famous quantum Zeno regime for realistic probe laser intensities.
 \end{abstract}

 \maketitle
\section{Introduction}
\label{sec:introduction}
The measurement of static or time-dependent quantities is a fundamental task in physics when studying the properties of a system. Thermal drift, imperfect detectors, or external noise sources often limit the accuracy of gathered data. Even if all these sources are absent, quantum mechanics has taught us that quantum measurements are inherently noisy. Outcomes of a measurement are in general stochastic and induce a back-action on the measured system that depends on the outcome of the measurement.
The very active field of quantum electronics provides many examples where the quantum properties of nanoscopic objects immediately lead to time-dependent fluctuating streams of measured
data that need to be analyzed by statistical methods \cite{ubbelohdeNATCOMM2012,gustavssonPRB2007}. John von Neumann started to develop a quantum theory of measurement based on projection operators \cite{neumannBOOK1955}. Later, Misra and Sudarshan found that repetitive projective measurements of the same quantity lead to the suppression of system dynamics, the celebrated quantum Zeno effect \cite{misraJMP1977}. This holds, however, only true in the case of strong measurements where any single measurement forces the system to immediately reveal its state with respect to the measurement operator. Aharonov and others then introduced a mathematical concept for describing so-called weak measurements that perturb the system only weakly on the expense of gaining only partial information \cite{aharonovPRL1988}. Korotkov studied theoretically a quantum dot system that is continuously monitored via a probe current that couples to the system with a tuneable coupling constant \cite{korotkovPRB2001}. 
The time-dependent measured current signal $z(t)$ was shown to contain a contribution from quantum Rabi-Oscillations that disappeared in the limit of strong measurements due to the quantum Zeno effect. 
The appearance of Rabi-oscillations in $z(t)$ demonstrates the usefulness of continuous weak measurements
to probe the dynamics of quantum systems. Since $z(t)$ is directly obtained from
experiment, we consider it the fundamental quantity in measurement theory. 
Starting in the 1980's a consistent theory of measurements in continuous
time (quantum continuous measurements) was established that was able to describe both the quantum system and the the stochastic detector output \cite{barchielliNC1982,belavkinConf1987}. In quantum trajectory theory, continuous measurements are treated with the help of stochastic differential equations \cite{barchielliBOOK2009,boutenSIAM2007}. The so-called stochastic master equation (SME) of a general quantum system and an arbitrary observable in the notation of Jacobs and Steck \cite{jacobsCP2006} will here be used to formulate a general theory of statistical properties of $z(t)$. 
 Equivalent SMEs have been derived by several authors in the context of special quantum systems 
\cite{belavkinConf1987, diosiPLA1988,gagenPRA1993,korotkovPRB1999,korotkovPRB2001,goanPRB2001}. The most general  derivation of the SME and an account of many of its properties has been given by Barchielli and Gregoratti in Ref. \onlinecite{barchielliBOOK2009} with an emphasis on rigorous derivations starting from a solid understanding of Ito stochastic calculus.  
They also present formulas for the usual second order noise power spectrum but do not evaluate their formulas with respect to moments or cumulants beyond second order. 
The SME is universal in the sense that it covers in principle also strong measurements as any strong measurement can be decomposed into a series of weak measurements \cite{oreshkovPRL2005}.
The stochastic detector output obtained from the SME was shown to be dominated by Gaussian shot noise for weak measurements
and to exhibit stochastic switching behavior (telegraph noise) for strong continuous measurements \cite{korotkovPRB2001}.  
 Starting from the multi-time moments of $z(t)$, we derive as the fundamental result of this work analytic expressions for cumulants that are correct in all orders of the measurement strength (Sections \ref{sec:Multitime} and \ref{sec:multicum}). 
 Despite an initially very unwieldy representation of the cumulants in terms of moments we can extremely simplify the cumulant expressions by introducing a slightly modified quantum propagator. Cumulants are of great importance for actual measurements as they allow for a straightforward subtraction of instrumental background noise which is not possible in the case of  moments (see Section \ref{sec:cumulants}) .   
 The new cumulant expressions are the basis for deriving quantum mechanical expressions with 
 unambiguous operator ordering  for higher order spectra  (Section \ref{sec:Trispectrum}). While a general theory of such spectra was elusive before, such spectra were found to give important additional information on quantum systems like internal interactions or correlation effects \cite{liPRL2016,ubbelohdeNATCOMM2012}. We will treat the non-trivial case of a strongly coupled spin pair and discuss its second, third and fourth order spectra.

{\it Path integrals} have alternatively been investigated to describe continuous quantum measurements \cite{cavesPRD1986,menskyPLA1994}. Bednorz {\it et al.} used such an approach to arrive at expressions for higher order correlation functions of $z(t)$ without deriving an explicit stochastic expression for $z(t)$ \cite{bednorzNJP2012}. 
Our derivation of  higher order correlation functions in Section \ref{sec:Multitime} is more general as we include the case of damping by the environment and - very importantly - can relax Bednorz'  assumption of a weak measurements. For absent damping we find complete agreement with Bednorz's result.  

{\it Langevin approaches} have been used in the past to formulate a theory of spin noise spectroscopy based on the fluctuating electron spin $s_z(t)$ \cite{glazovPRB2012,liPRL2016}. These approaches, however, do not regard the effects of measurement back-action on the spin system and can not be used to derive expressions for higher order noise spectra (Section \ref{sec:Langevin}).

 In 1959, {\it Landau} was probably the first to popularize in his book on Statistical Physics a simple quantum theory for the noise spectrum of an observable given by a Hermitian operator $A$ by considering the noise spectrum of a classical quantity \cite{landauBOOKstat}.
 Earlier versions have e.g. appeared in context with the fluctuation dissipation theorem \cite{kuboJPSJ1957}.
The classical noise power spectrum of a fluctuating quantity $z(t)$ is defined via the autocorrelation of $z(t)$ as
\begin{equation}
 S_z(\omega) = \int_{-\infty}^\infty e^{i\omega \tau}\langle z(t+\tau) z(t) \rangle_t  \,{\rm d}\tau, 
 \label{ClassicalNoiseSpectrum}
\end{equation}
where the brackets $\langle \cdots \rangle_t$ represent an average over time $t$.
Equivalently, the noise spectrum can be expressed in terms of the Fourier transform $z(\omega)$ 
of $z(t)$ as  \cite{gardinerBOOK2009} (see Appendix \ref{app:Conventions} for definition of the Fourier transform) 
\begin{equation}
  \langle z(\omega)z^*(\omega')\rangle = 2 \pi \delta(\omega - \omega') S_z(\omega)  
\end{equation}
where  $z^*(\omega')$ denotes the complex conjugate of $z(\omega')$.

Landau defines without derivation  \cite{landauBOOKstat}
\begin{equation}
 S^{\rm (L)}(\omega) = \int_{-\infty}^{\infty} e^{i\omega \tau} G^{\rm (L)}(\tau) \,{\rm d}\tau \label{LandauFormula1}
\end{equation} 
with [the superscript (L) refers to Landau]
\begin{equation}
  G^{\rm (L)}(t'-t) = \frac{1}{2}{\rm Tr}([A(t) A(t') + A(t') A(t)]\rho_0) \label{LandauFormula2}
\end{equation}
as the quantum mechanical analog of Eq.~(\ref{ClassicalNoiseSpectrum}) where the product of the classically measured quantity $z(t)$ has been replaced by a symmetrized product of the measured operator $A$ \cite{landauBOOKstat}. The Heisenberg picture is invoked to describe the time dependent operator $A(t)$. The density matrix of the system $\rho_0$ is constant in time and assumed to be in steady state  from which follows that the RHS of Eq.~(\ref{LandauFormula2}) depends only on the temporal difference $t'-t$.  Landau's formula can in principle be used to describe damping if coupling of the quantum system to a bath is included. This can be done by replacing $\rho_0$ by a density matrix $\chi_0$ which represents system and bath. A useful expressions for the reduced density matrix $\rho_0$ is then obtained via the quantum regression theorem where damping is treated in Markov approximation (see Section \ref{sec:QRT}). We show that the same formula appears also naturally from the SME (see Section \ref{sec:NoisePower}).

In the following sections, we will choose spin noise spectroscopy as an illustrative example along which we will identify important quantities of our theory in the experimental setup. 
 Spin noise spectroscopy has after pioneering work by Aleksandrov and Crooker \cite{aleksandrovJETP1981,crookerNATURE2004} quickly evolved into a mighty tool for studying spin systems in semiconductors and gases \cite{oestreichPRL2005,mullerPHYSICAE2010,aleksandrovJP2011,hubnerPSSB2014,glazovJETP2015,sinitsynRPP2016}.
 \begin{figure}
   \centering
 \includegraphics[width=8cm]{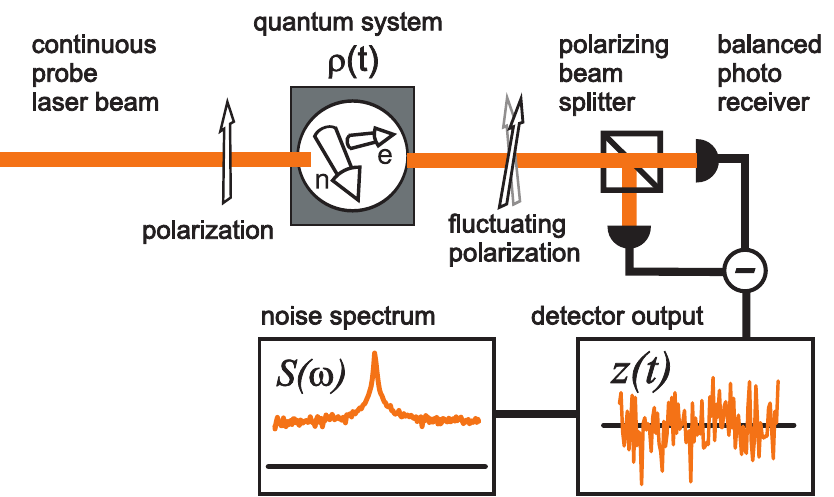} 
   \caption{Schematics of continuous spin noise spectroscopy. The fluctuating electron spin orientation in a quantum system induces a fluctuating polarization of the probe beam via the Faraday-effect. Polarization noise and photon shot noise contribute to the noise signal $z(t)$. The noise spectrum $S(\omega)$ exhibits a peak at the spin precession frequency and a constant offset due to shot noise. The system $\rho$ may include coupling of the electron spin (e) with a nuclear spin (n).  }
    \label{spinNoiseSchematics}
    \end{figure}
The spin noise setup shown in Fig. \ref{spinNoiseSchematics} realizes a continuous quantum measurement of the $z$-spin orientation of an electron via the Faraday-effect in a semiconductor sample. Oestreich {\it et al.} give an estimate for the fluctuating Faraday angle based on the Elliott-formula which is based on band-structure parameters \cite{oestreichPRL2005}. The electron may in general be part of a larger coupled quantum system like e.g. an interacting pair of an electron spin and a nuclear spin \cite{bussPRB2016}. A finite Faraday-angle of the probe light polarization after the sample leads after a polarizing beam splitter to a slight imbalance of light intensities on the photodiodes. The Faraday signal is superimposed by strong optical shot noise. The $z$-spin orientation in the quantum system is consequently not instantly revealed by the Faraday effect as expected for a weak quantum measurement. Spin precession frequencies and spin lifetimes of electron spins can be deduced from the peak position and peak width of the noise power spectrum \cite{oestreichPRL2005}.

Several recent quantum theories for calculating spin noise spectra used Landau's Eq.~(\ref{LandauFormula1}) 
without including a bath and are therefore missing a possible treatment of spin relaxation via the quantum regression theorem \cite{braunPRB2007,clerkRMP2010,sinitsynRPP2016}. 
Alternatively, semiclassical Langevin equations of motion for single spin systems or electron-hole spin systems were used that included damping \cite{glazovPRB2012,smirnovPRB2014}.
 While Korotkov already elaborates on the effect of measurement back-action for the case of a nano-electronic device \cite{korotkovPRB2001}, most of the spin noise theories do not include the effect of measurement induced damping into the calculation of spin noise spectra. An exception is Ref. \onlinecite{smirnovPRB2017} where measurement back action is treated numerically for a special quantum dot system. We show in Section \ref{sec:Zeno} that this damping leads to an additional anisotropic spin dephasing whose strength can easily be estimated from experiment and eventually induces Zeno-physics for high probe laser intensities. 
 
 Recently, higher order noise spectra of $z(t)$ came in the focus of interest for obtaining additional information on quantum systems. Liu showed theoretically that third order moments of the noise signal can be used to distinguish homogenous from inhomogeneous broadening of the spin noise resonance \cite{liuNJP2010} providing an alternative to a recently introduced method which relies on a wavelength scan of the probe laser \cite{zapasskiiPRL2013}. 
 The (third order) bi\-spectrum of the current through a single electron device was measured by Ubbelohde and shown to reveal correlation effects \cite{ubbelohdeNATCOMM2012}.
 First steps into developing a theoretical understanding of a fourth order spin noise correlation spectrum
\begin{equation}
 S^{\rm (corr)}(\omega_1,\omega_2) = \langle |z(\omega_1)|^2 |z(\omega_2)|^2 \rangle - \langle |z(\omega_1)|^2 \rangle \langle|z(\omega_2)|^2 \rangle \label{eq:fourthOrderCorrelation}
\end{equation}
 were undertaken by Li and Sinitsyn \cite{liNJP2013,liPRL2016,liPRA2016}. Practical broadband real-time measurements of $S^{\rm (corr)}(\omega_1,\omega_2)$ were presented by Starosielec {\it et al.} who were able to measure two-dimensional spectra up to 90~MHz with a resolution of better than 1~MHz in real time \cite{starosielecRSI2010}. 
 We will show in Section \ref{sec:S4spinpair} that higher order spectra $S_z^{(n)}$ not only yield   information on inhomogeneous broadening but also reveal correlations of coherently coupled quantum systems like an electron-nuclear spin pair \cite{bussPRB2016}.
 
The paper is organized as follows: Sections \ref{sec:cumulants} and  
 \ref{sec:SME} are used to shortly  review some important properties of cumulants, higher order spectra, and the so-called (non-linear) stochastic master equation. An iterative solution of the SME in terms of Ito-Integrals yields then explicit expressions for $z(t)$ in ascending orders of the measurement strength (Section \ref{sec:CQNF}). A quantum mechanical expression of the usual power spectrum of $z(t)$  is obtained in terms of the quantum propagator (Section \ref{sec:NoisePower}) and compared with Landau's approach in context with the quantum regression theorem and the fluctuation dissipation theorem (Sections \ref{sec:LandauApproach}, \ref{sec:QRT} and Appendix \ref{app:FDT}). In Section \ref{sec:Zeno} we show that the ratio of shot noise background and the surface area under noise peaks in a measured spectrum contains important information on the measurement strength. Based on this result we argue that the Zeno effect might have been observed already in a recent spin noise experiment.  
In Section \ref{sec:Examples} we present analytical as well as numerical examples for the calculation of spin noise spectra including the non-trivial case of a strongly coupled pair of spins. 
In Section \ref{sec:Langevin} we shortly compare an earlier theory of spin noise spectroscopy that is based on a Langevin-equation with our approach based on the SME. 
Section \ref{sec:Multitime} presents quantum mechanical expression for multi-time moments of the detector output $z(t)$ that are directly derived from the SME without any approximations. The derivation does not rely on explicit expressions for $z(t)$ as derived in 
Section \ref{sec:CQNF} and is also more general than a previously found derivation by Bednorz based on path-integrals \cite{bednorzNJP2012}. A reader who is especially interested in beyond second order results may skip Sections \ref{sec:CQNF} to  \ref{sec:SignalToNoise}. These sections, however, are necessary for understanding the connection of  the higher order spectra $S_z^{(n)}$ with the different orders $z_j(t)$ of the detector output. 
The expressions for the multi-time moments are the starting point in Section \ref{sec:multicum} for finding compact expressions for the third and  fourth order multi-time cumulants whose structure,  intriguingly, very much resembles those of the moments. 
After Fourier transformation of the cumulants we eventually obtain the bispectrum and trispectrum (Section \ref{sec:Trispectrum}). In the last Section \ref{sec:S4spinpair} we present the numerics of higher order noise spectra of a coupled electron-nuclear spin pair. Table \ref{TableSymbols} displays the most important symbols with references to the relevant equations.
\begin{table*}
\caption{\label{TableSymbols} List of the most important symbols used in the text.}
 \begin{ruledtabular}
\begin{tabular}{l l}
Symbol & Meaning \\
\hline
$\rho(t)$ & system density matrix \\
$\rho_0$ & steady state density matrix \\
$H$ & Hamilton operator \\
$A$ & measurement operator \\
$\beta$ & measurement strength \\
${\cal D}$ & superoperator of damping by environment (in Markov approximation) \\
${\cal L}$ & Liouville superoperator ${\cal L}\rho = i[\rho,H]/\hbar + {\cal D}\rho - \beta^2[A,[A,\rho]]/2$ \\
& including damping by environment and measurement \\
${\cal G}(t)$, ${\cal G}'(t)$ & propagator ${\cal G}(t) = e^{{\cal L }t}$  for $t>0$ and modified propagator ${\cal G}'(t) = {\cal G}(t) - {\cal G}(\infty)\Theta(t)$ \\
$\Theta(t)$ & Step function with $\Theta(t) = 1$ for $t>0$ and zero otherwise. \\ 
${\cal A}$, ${\cal A}'$ & measurement superoperator ${\cal A} x = (Ax + xA)/2$ and modified operator ${\cal A}' = {\cal A} - {\rm Tr}( A \rho_0)$\\
${\rm d}W$ & infinitesimal increment of a Wiener process (Ito calculus) \\ 
$\Gamma(t)$ & white noise, formal derivate of ${\rm d}W/{\rm d}t$\\
$z(t)$ & detector output $z(t) = \beta^2 {\rm Tr}(A \rho(t)) + \beta \Gamma(t)/2$\\
$G_{\rm q}(\tau)$ & cumulant based quantum noise autocorrelation $G_{\rm q}(\tau) = \frac{1}{2}\langle A(|\tau|) A(0) + A(0) A(|\tau|) \rangle -  \langle A(0) \rangle^2$ \\
& in Heisenberg picture with possible damping of $A$ \\
$S_{\rm q}(\omega)$ & quantum noise power spectrum without shot noise background  $S_{\rm q}(\omega) = \int_{-\infty}^\infty  G_{\rm q}(\tau) e^{-i \omega \tau} \,{\rm d} \tau$ \\
$C_n(x_1,\cdots, x_n)$ & $n$th order cumulant of stochastic vector components $x_j$, see Eq.~(\ref{eq:cumulantgeneration})\\
$M_n$ & multi-time moments $\langle z(t_n)\cdots z(t_1) \rangle$, quantum expressions see Eq.~(\ref{eq:MomentsK}) \\
$C_n( z(t_n),\cdots, z(t_1))$ & multi-time cumulants, quantum expressions for $n=3$ and $n=4$ see Eqs.~(\ref{eq:C3}) and (\ref{eq:C4}) \\ 
$S_z^{(n)}$ & cumulant based $n$th order polyspectrum of fluctuating quantity $z(t)$ defined in Eq.~(\ref{DefPolyspectra}) \\
$S_z^{(2)}(\omega)$ & powerspectrum of $z(t)$  where $S_z^{(2)}(\omega) = \beta^4 S_{\rm q}(\omega) + \beta^2/4$,
see Eqs.~(\ref{O6S}) and (\ref{eq:S2})
\\
$S_z^{(3)}(\omega_1,\omega_2)$ & bispectrum of $z(t)$, see Eq.~(\ref{eq:S3})\\
$S_z^{(4)}(\omega_1,\omega_2,\omega_3)$ & trispectrum of $z(t)$, see Eq.~(\ref{eq:S4})
\end{tabular}
 \end{ruledtabular}
\end{table*}

\section{Cumulants and higher order spectra} 
\label{sec:cumulants}
Fluctuating quantities like $z(t)$ need to be characterized by quantities such as the average mean $\langle z(t) \rangle$ or the covariance $\langle z(t+\tau) z(t)  \rangle_t -  \langle z(t+\tau) \rangle_t \langle z(t) \rangle_t$. A generalization of such quantities to higher orders is given by the so-called cumulants. 
The cumulants $C_n$ are defined via the cumulant generating function \cite{gardinerBOOK2009}
\begin{equation}
 K_{\vec{x}}(\vec{k}) = \ln\left\langle \exp(\vec{k}\cdot\vec{x})\right\rangle
\end{equation}
and its derivatives at $\vec{k} = 0$ by
\begin{equation}
 C_n(x_1,\cdots,x_n)  =   \frac{\partial^n}{\partial k_1\cdots\partial k_n} K_{\vec{x}}(\vec{k}) |_{\vec{k}= 0}, 
 \label{eq:cumulantgeneration}
\end{equation}
where $x_j$ are the components of a stochastic vector $\vec{x}$. The expressions
\begin{eqnarray}
 C_1(x_1) & = & \langle x_1 \rangle \nonumber \\
 C_2(x_1,x_2) & = & \langle x_1 x_2 \rangle - \langle x_1 \rangle\langle x_2 \rangle\nonumber \\
 C_3(x_1,x_2,x_3) & = & \langle x_1 x_2 x_3 \rangle - \langle x_1 x_2 \rangle \langle x_3 \rangle \nonumber\\
 &  & \hspace{-2cm}-  \langle x_1 x_3 \rangle \langle x_2 \rangle  - \langle x_2 x_3 \rangle \langle x_1 \rangle + 2   \langle x_1 \rangle \langle x_2 \rangle \langle x_3 \rangle \label{eq:cumulants}
\end{eqnarray} 
follow, where we identify $C_1$ as the average mean and $C_2$ as the covariance.
The variance of $x_1$ is obtained from $C_2$ for $x_2 = x_1$. The expression for $C_4$ can be found in
Appendix \ref{AppC4}, Eq.~(\ref{eq:Cumulant4}).
For the sum of two {\it independent} stochastic vectors $\vec{x}$ and $\vec{y}$
we find  $K_{\vec{x}+\vec{y}}(\vec{k}) = K_{\vec{x}}(\vec{k}) +K_{\vec{y}}(\vec{k})$ because of $\ln(\langle a b \rangle) = \ln(\langle a  \rangle) + \ln( \langle  b \rangle)$ for independent positive stochastic quantities $a$ and $b$. 
Consequently, any cumulant shares the important property of linearity  \cite{nikiasIEEE1993} 
\begin{equation}
 C(x_1+y_1, x_2 + y_2,...) = C(x_1,x_2,...) + C(y_1,y_2,...) \label{cumulantsLin}
\end{equation}
if $\vec{x}$ and $\vec{y}$ are independent stochastic quantities.
Any definition of a noise spectrum is therefore required to be a cumulant in order to guarantee that e.g. the spectrum of background noise $z_{\rm b}(t)$ can be subtracted from the spectrum of the desired quantity.  Any ordinary moment of $z(t) + z_{\rm b}(t)$ would suffer from spurious cross-contributions of $z(t)$ and $z_{\rm b}(t)$.   

The traditional definition of the noise spectrum $S_z(\omega)$ of $z(t)$ according to Eq.~(\ref{ClassicalNoiseSpectrum}) is 
a cumulant only for the case of $\langle z(t) \rangle = 0$ which holds if $z(t)$ was defined accordingly. Instead, we use here a second order spectrum $S_z^{\rm (2)}$ that is directly based on the cumulant $C_2$ 
\begin{equation}
 C_2(z(\omega_1),z(\omega_2)) = S_z^{\rm (2)}(\omega_1) 2\pi \delta(\omega_1+\omega_2). \label{eq:powerspectrum}
\end{equation} 
where $\delta$ is the Dirac delta function.
It can be easily verified that $S_{z}^{(2)}(\omega)$ of $z$ is related to $S_z(\omega)$ via
\begin{equation}
 S_{z}^{(2)}(\omega) = S_{z-\langle z\rangle}(\omega),
\end{equation}
i.e. the two definitions agree for fluctuating quantities that are average-free. In the following we will always use $S_z^{\rm (2)}(\omega)$, because the quantum expressions for $z(t)$ that we are going to derive are in general not average-free.

The generalization to the higher order spectra $S^{(n)}$ (so-called polyspectra) had been given by Brillinger in 1965 as \cite{brillingerAMS1965,ubbelohdeNATCOMM2012}
\begin{eqnarray}
C_n(z(\omega_1),\cdots, z(\omega_n))  & = &    S_z^{(n)}(\omega_1,\cdots,\omega_{n-1}) \nonumber \\
 & & \hspace{-2cm} \times 2\pi \delta(\omega_1 +\cdots+ \omega_n). \label{DefPolyspectra}
\end{eqnarray}
The third and fourth order polyspectra  $S_z^{(3)}$ and $S_z^{(4)}$ are often referred to as the bispectrum and trispectrum, respectively
\cite{nikiasIEEE1993}. The bispectrum has been used in different fields of physics to e.g. reveal time-asymmetries in geologic behavior \cite{efimovQJRMS2001}, to investigate magnetization fluctuations \cite{balkPRX2018}, or to investigate current fluctuations in nanoscale structures. Ubbelohde {\it et al.} measured the bispectrum of a tunneling current through a single electron transistor \cite{ubbelohdeNATCOMM2012}. The appearance of a clear non-zero bispectrum gave evidence for the non-Gaussian temporal fluctuations of the current. Temporal traces of the current showed telegraph noise indicating the limit of a strong measurement. 
 Since cumulants $C_n$ with $n \geq 3$ are strictly zero for Gaussian behavior \cite{gardinerBOOK2009}, the bi- and trispectum are an indicator for non-Gaussian fluctuations.
 \begin{widetext}
 Last we show that $S_z^{(n)}$ is free from $\delta$-function contributions under very general conditions.
  In contrast $S_z$ [Eq.~(\ref{ClassicalNoiseSpectrum})] contains for $\langle z(t) \rangle \neq 0$ always a contribution $2\pi \delta(\omega) \langle z(t) \rangle^2$.
 Consider the multi-time cumulant for $n=3$ 
 \begin{eqnarray}
 C_3(z(t_1 = \tau_0),z(t_2 = \tau_0 + \tau_1),z(t_3 = \tau_0 + \tau_1 + \tau_2)) & = & \frac{\partial^3}{\partial k_1 \partial k_2\partial k_3} \ln\left\langle \exp(k_1 z(t_1) +k_2 z(t_2) + k_3 z(t_3))\right\rangle.
 \end{eqnarray} 
 This expression is for stationary processes $z(t)$ independent of $\tau_0$. Consequently, the Fourier transformation with respect to $\tau_0$ results in the $\delta$-function of Eq.~(\ref{DefPolyspectra}). For large $\tau_2$ we note that $z(t_1)$ and $z(t_2)$ become increasingly independent of $z(t_3)$. The cumulant $C_3$ assumes zero for $\tau_2 \rightarrow \pm \infty$ since
 \begin{eqnarray}
   C_3 & = & \frac{\partial^3}{\partial k_1 \partial k_2\partial k_3} \ln\left\langle \exp(k_1 z(t_1) +k_2 z(t_2) + k_3 z(t_3))\right\rangle \nonumber \\
    & = & \frac{\partial^3}{\partial k_1 \partial k_2\partial k_3} \ln\left\langle \exp(k_1 z(t_1) +k_2 z(t_2)) \rangle\langle \exp( k_3 z(t_3))\right\rangle \nonumber \\
    & = & \frac{\partial^3}{\partial k_1 \partial k_2\partial k_3}\left( \ln\left\langle \exp(k_1 z(t_1) +k_2 z(t_2))\right\rangle
    + \ln \left\langle \exp( k_3 z(t_3))\right\rangle \right)\nonumber \\
    & = & 0.
 \end{eqnarray}
 The second line regards that averages of products of independent processes can be written as products of their averages. The last but one line disappears after the partial derivates since the first term does not depend on $k_3$ and the last term does not dependent on $k_1$ and $k_2$. Similarly we find $C_3 = 0$ for $\tau_1 \rightarrow \pm \infty$. A Fourier transformation of $C_3$ with respect to $\tau_j$ except $\tau_0$ therefore never exhibits a $\delta$-function contribution which shows that $S_z^{(3)}$ is free from $\delta$-function contributions. Using corresponding arguments it can be shown that all polyspectra $S_z^{(n)}$ are free from $\delta$-function contributions. 
 We will make use of this finding in Section \ref{sec:Trispectrum}.
 \end{widetext}

\section{stochastic master equation}
\label{sec:SME}
In this section, we shortly review the stochastic master equation (SME) that governs the time dependent density matrix $\rho(t)$ of a quantum system that is continuously monitored for an observable $A$ (hermitian operator) yielding a detector output $z(t)$. In 1987, Belavkin was probably the first to publish a version of the SME \cite{belavkinConf1987}. Corresponding equations were derived several times independently by Diosi, Gagen, Korotkov, and Goan
 for special cases in transport theory 
 and quantum optics \cite{diosiPLA1988,gagenPRA1993,korotkovPRB1999,goanPRB2001}. Especially, Diosi gives a thorough step by step derivation of the SME  for continuous position measurement on a particle. A very general derivation and treatment of the SME with an emphasis on mathematical rigor is given by two pioneers of the field, Barchielli and Gregoratti \cite{barchielliBOOK2009}.
A recent review by Jacobs and Steck gives an easy to follow derivation of the SME in terms of so-called 'positive operator-valued measures' (POVMs) \cite{jacobsCP2006}.
After a measurement, the density matrix assumes the form
\begin{equation}
  \rho_{\rm f} = \frac{\Omega_m \rho \Omega_m^\dagger}{{\rm Tr}[\Omega_m \rho \Omega^\dagger_m]},
  \label{POVM}
\end{equation}
with the probability $P(m) = {\rm Tr}[\Omega_m \rho \Omega^\dagger_m]$ of finding the detector value $m$.
The operators $\Omega_m$ need to fulfill the relation $\sum_m \Omega_m \Omega^\dagger_m = \mathbbm{1}$.
Strong measurements, $\Omega_m = |m\rangle \langle m|$, cause $\rho$ to collapse into an eigenstate $ |m\rangle$ of the observable $A$. Weak measurements can be modelled by operators $\Omega_m$ that are mixtures of 
 $|m\rangle \langle m|$  and tend towards $\Omega_m \propto \mathbbm{1}$ for very weak and eventually absent measurements. Annabestani {\it et al.} analyzed in some detail the distribution and joint distribution of measurement results after one and two subsequent weak measurements, respectively, for the special case of an ensemble of electron spins in an ESR experiment \cite{annabestaniJMR2015}. They, however, did not proceed to derive the case of a continuous measurement which should result in the SME for their quantum system. 
 The SME for a general $\rho(t)$ appears by Jacobs after considering the statistical properties of $P(m)$ from which he finds that the detector output $z(t)$ is basically Gaussian noise with the center of the Gaussian shifted proportional to the expectation value
 ${\rm Tr}(A \rho(t))$. Consequently, the detector output is the expectation value ${\rm Tr}(A \rho(t))$ hidden behind a strong background noise, which, however, is expected for a weak measurement that is unable to fully reveal the result for a
 measurement of $A$.
The stochastic master equation 
in the Schr\"odinger picture (Ito-calculus)
\begin{eqnarray}
    {\rm d}\rho & = & \frac{i}{\hbar}[\rho,H]{\rm d}t + {\cal D}\rho \,{\rm d}t \nonumber \\
     & & - \frac{\beta^2}{2}[A,[A,\rho]] {\rm d}t\nonumber \\
     & & + \lambda \beta \left[A\rho + \rho A  - 2 \rho {\rm Tr}(\rho A) \right]{\rm d}W \label{SME} 
\end{eqnarray}        
propagates the density matrix $\rho(t)$ of a quantum system which is continuously monitored for the expectation value of an operator $A$ \cite{jacobsCP2006}. The measurement strength is quantified by $\beta$. The quantity ${\rm d}W$ is a zero-mean Gaussian random variable with variance ${\rm d}t$, i.e. formally ${\rm d}W$ scales as $\sqrt{{\rm d}t}$ \cite{gillespieAJP1995}. It is this quantity through which a correct description of the randomness in quantum measurements appears in the theory. Ito-calculus uses the non-anticipating differential  ${\rm d}\rho = \rho(t + {\rm d}t) - \rho(t)$ whereas Stratonovich calculus uses  
${\rm d}\rho = \rho(t + {\rm d}t/2) - \rho(t - {\rm d}t/2)$ \cite{gardinerBOOK2009}. The parameter $\lambda$ will below be used for a successive approximation of the solution for $\rho(t)$. The usual SME is given for $\lambda = 1$. 
A second stochastic equation 
\begin{equation}
      {\rm d}Z  =  \beta^2 {\rm Tr}(\rho A){\rm d}t +\beta \frac{1}{2} {\rm d}W \label{SME_detector}  
\end{equation}
describes the evolution of the time-integrated detector output $Z(t)$.  
The formal derivative $z(t) = \dot{Z}(t)$
yields the time dependent detector output 
\begin{equation}
      z(t)  =  \beta^2 {\rm Tr}(\rho(t) A) +\beta \frac{1}{2} \Gamma(t) \label{SME_detector}  
\end{equation}
where $\Gamma(t)$ with $\langle \Gamma(t) \Gamma(t') \rangle = \delta(t-t')$ represents white Gaussian noise. 
  The first line of Eq. (\ref{SME}) is equivalent to the usual master equation of a unobserved quantum system. The Hamilton operator $H$ describes the coherent evolutions of the system. A linear superoperator ${\cal D}$ models damping of the quantum system due to coupling with its environment in Markov approximation. The effect of a continuous measurement of operator $A$ on $\rho(t)$ is modelled by the second and third line. Continuous measurement leads to a gradual decay of the system towards an eigenstate of $A$  that is described by a damping term proportional to $\beta^2$ in the second line (see also Ref. \onlinecite{bednorzNJP2012}). Within continuous measurement theory this term appears to be independent from the actual realization of the continuous measurement \cite{jacobsCP2006}. 
 In an actual spin noise experiment $\beta^2$ scales with the probe laser intensity [compare Eq.~(\ref{O6S})]. The laser induces an additional anisotropic spin dephasing with dephasing rates
  $\gamma^{\rm M}_x =  \gamma^{\rm M}_y = 2 \beta^2$ and   $\gamma^{\rm M}_z = 0$ (see Section \ref{sec:Zeno}). 
The quantum Zeno effect and telegraph noise behavior of $z(t)$ follow from Eq.~(\ref{SME}) in the case of strong coupling ($\beta^2 \gg 1$), as shown previously for the special cases of a continuous position measurement \cite{gagenPRA1993} and
 a two level system \cite{korotkovPRB2001}. The last line of the SME establishes a stochastic back-action of the measurement result on $\rho$. Please note that the last line is non-linear in $\rho$. The appearance of such a non-linearity is no surprise since any quantum measurement destroys quantum linearity [see Eq.~(\ref{POVM})].
 The  detector output [Eq.~(\ref{SME_detector})] consists of the expectation value of $A$ and Gaussian noise $\Gamma(t)$.  Gaussian noise correctly represents shot noise as long as - in the case of spin noise spectroscopy - enough photons contribute to the signal within a temporal interval given by the detector's temporal resolution. 
 
 Concluding the section, we want to emphasize that (i) the SME has the same form independent of the actual experimental realization of a continuous measurement as long as a measurement of the observable $A$ is realized; Similarly, no specifics of the detector need to enter the SME.
 (ii) the SME includes the case of strong measurements for large $\beta$ and is able to reproduce switching behavior (telegraph noise) of the detector output that corresponds to collapses of the system into eigenstates of $A$. 
 We therefore consider the SME as the approach of choice for treating quantum noise of general quantum systems and arbitrary measurement strength.
\section{Expressions for detector output $z(t)$:
The continuous quantum noise formulas} 
\label{sec:CQNF}
The above SME is our starting point for calculating the fluctuating detector output $z(t) = \dot{Z}(t)$ in orders of $\beta$ from an iteration scheme. The first three orders turn out to have clear physical interpretations and are sufficient to calculate the second order spectrum $S_z^{(2)}(\omega)$. The derivation of third and fourth order spectra starts in Section \ref{sec:Multitime} with the calculation of multi-time moments. 

The first two lines of Eq.~(\ref{SME}) will be abbreviated with ${\cal L}\rho$ where
${\cal L}$ is a linear superoperator, the so-called Liouvillian, which describes the coherent and incoherent evolution of $\rho$ without the stochastic back-action.
%We formally introduce the temporal derivative $\Gamma(t) = {\rm d}W/{\rm d}t$ which represents white Gaussian noise 
%(Langevin noise) with the important property $\langle \Gamma(t)\Gamma(t') \rangle = \delta(t-t')$. 
The rewritten SME
\begin{equation}
  {\rm d}{\rho} = {\cal L}\rho\, {\rm d}t + \lambda \beta [A  \rho + \rho A  - 2 \rho {\rm Tr}(\rho A)]\,{\rm d}W
  \label{SME2}
\end{equation}  
can be solved by the method of successive approximation using the ansatz
\begin{equation}
 \rho = \rho_0 + \lambda \rho_1 + \lambda^2 \rho_2 + ... \label{successiveAnsatz}
\end{equation} 
which  gives by comparing orders of $\lambda$
 \begin{equation}
  {\rm d}\rho_0 = {\cal L}\rho_0 \,{\rm d}t  \label{SME_zero}
 \end{equation}
and
 \begin{equation}
     {\rm d}\rho_1 = {\cal L}\rho_1\, {\rm d}t + \lambda \beta [A  \rho_0 + \rho_0 A  - 2 \rho_0 {\rm Tr}(\rho_0 A)]\,{\rm d}W
    \label{SME_first}
 \end{equation}
 for the zeroth and first order contributions.
 The $n+1$-order contribution is
 \begin{equation}
      {\rm d}\rho_{n+1} = {\cal L}\rho_{n+1}\, {\rm d}t + \lambda \beta B_n(\rho_n,\rho_{n-1},\cdots,\rho_0)\,{\rm d}W,
      \label{SME_approx}
 \end{equation}
where we define
\begin{equation}
  B_n(\rho_n,\cdots,\rho_0) = 2{\cal A}\rho_n - 2 \sum_{\nu = 0}^n \rho_\nu {\rm Tr}(A \rho_{n-\nu})
\end{equation}
with the superoperator ${\cal A}x = (A x + x A)/2$.

 The zeroth order contribution $\rho_0(t)$ reaches a constant equilibrium (steady state) $\rho_0$ for $t \rightarrow \infty$ due to damping. We  therefore identify $\rho_0$ as the equilibrium state of the continuously monitored quantum system. Please note that 
an increasing measurement strength $\beta$ drives $\rho_0$ away from the true thermal equilibrium due to the 
measurement induced damping.
 Eqs.~(\ref{SME_approx}) can be interpreted as first order linear differential equations for $\rho_{n+1}$ with driving terms that depend on $\rho_n$. They are conveniently solved after introducing a Greens function-like superoperator ${\cal G}(t) = \exp({\cal L}t)$ for $t > 0$ and with ${\cal G}(t) = 0$ for $t \le 0$. Using the star $\star$ for a convolution 
 obeying Ito-calculus 
 \begin{equation}
  {\cal G}(t) \star a(t) = \int_{-\infty}^{\infty} {\cal G}(t-\tau)a(\tau) \,{\rm d}W(\tau) \label{ItoConv2}
\end{equation}
 (see also Appendix \ref{sec:Ito})
we obtain 
\begin{equation}
  \rho_{n+1}(t)  = \beta {\cal G}(t) \star B_n(\rho_n,\cdots,\rho_0)
\end{equation}
or explicitly 
\begin{eqnarray}
  \rho_1(t) & = & \beta {\cal G}(t) \star B_0(\rho_0), \\
  \rho_2(t) & = & \beta^2 {\cal G}(t) \star B_1(  {\cal G}(t)  \star B_0(\rho_0), \rho_0).
\end{eqnarray}
The detector output $z(t) = z_1(t) + z_2(t) + z_3(t) + \cdots$ in orders of $\beta$ follows as
\begin{eqnarray}
 z_1(t) & = & \frac{\beta}{2} \Gamma(t) \nonumber \\
 z_2(t) & = & \beta^2 {\rm Tr}(A \rho_0)  \nonumber \\
  z_3(t) & = & \beta^3 {\rm Tr}[A {\cal G}(t) \star B_0(\rho_0) ) \nonumber \\
    & = & \beta^3 {\rm Tr}(A {\cal G}(t) \star (A  \rho_0 + \rho_0 A - 2 \rho_0  {\rm Tr}(A \rho_0) ) ]\nonumber \\
 z_4(t) & = & \beta^4 {\rm Tr}[A {\cal G}(t) \star B_1({\cal G}(t) \star B_0(\rho_0),\rho_0))]
\end{eqnarray}
and iteratively for $n+1$ as
\begin{equation}
 z_{n+1}(t)  =  \beta {\rm Tr}[A  {\cal G}(t) \star B_{n-2}(\rho_{n-2}(t),\cdots,\rho_0)].\label{eq:CQNF}
\end{equation} 
The expressions for $\rho(t)$ and $z(t)$ are correct in {\it all} orders of the measurement strength $\beta$. We will refer to the above equations as the continuous quantum noise formulas (CQNFs). Please note that the equations for $\rho(t)$ dependent non-linearly on $\rho_0$ as expected for a quantum system subject to measurements [compare Eq.~(\ref{POVM})].
The CQNFs  for $z(t)$
have well defined meanings in the case of a spin noise experiment.
The pure white noise contribution $z_1(t)$ is due to optical laser shot noise.  
A non-vanishing average $z$-spin orientation in thermal equilibrium leads to a constant Faraday rotation and therefore to a constant offset $z_2(t)$. 
The leading order contribution to noise from the actual quantum system is given by $z_3(t)$.
We emphasize that the appearance of $z_1$, $z_2$, and $z_3$ is generic and independent of the actual quantum system, or the way the weak continuous measurement is realized. In case of a transport experiment where a probe current is weakly coupled to the quantum system, $z_1$ would correspond to electronic shot noise, $z_2$ to the average probe current, and $z_3$ to the leading order influence of the system on the probe current dynamics.
Starting from the next section, we will use the expressions for $z_1(t)$, $z_2(t)$, and $z_3(t)$ to calculate the second order noise spectrum (the usual power spectrum) and compare it with expressions used in literature. Spectra beyond second order can in principle be calculated from the CQNF where, however, the number of intermediate terms in the calculation quickly rises. Alternatively, we will derive such expressions from multi-time moments of $z(t)$ 
that we calculate without any approximation in Section \ref{sec:Multitime} directly from considering the SME. No explicit expression for $z(t)$ will be required.

\section{The power spectrum}
\label{sec:NoisePower}
The power spectrum $S_z^{(2)}$ up to fourth order in $\beta$ follows from the Fourier transformation of the second order cumulant
\begin{eqnarray}
 C_2(z(t_1),z(t_2)) &  = & \langle z_1(t_1) z_1(t_2)\rangle + \langle z_3(t_1) z_1(t_2)\rangle  \nonumber \\
  & & + \langle z_1(t_1) z_3(t_2)\rangle + O(\beta^6), \label{C2O6}
\end{eqnarray}
where the constant offset $z_2$ disappears due to the properties of the cumulants.
We find
\begin{eqnarray}
 \langle z_1(t_1) z_1(t_2)\rangle & = & \frac{\beta^2}{4}\delta(t_1 - t_2) \nonumber \\
  \langle z_3(t_1) z_1(t_2)\rangle & = & \frac{\beta^4}{2} \int {\rm Tr}(A {\cal}G(t_1 - \tau)B_0(\rho_0)\langle\Gamma(\tau) \Gamma(t_2)\rangle {\rm d}\tau \nonumber \\
  & = & \frac{\beta^4}{2} {\rm Tr}(A {\cal G}(t_1-t_2)B_0(\rho_0)).
\end{eqnarray}
The cumulant $C_2$ depends only on the time difference $\tau = t_1 - t_2$. Consequently we can define a function
\begin{eqnarray}
  G(\tau) & = & C_2(z(t_1),z(t_2)) \nonumber \\
   & = & \frac{\beta^2}{4}\delta(\tau) + \beta^4 G_{\rm q}(\tau) + O(\beta^6) \label{O6G}
    \end{eqnarray}  
 where
 \begin{eqnarray}
  2 G_{\rm q}(\tau) & = & {\rm Tr}(A {\cal G}(\tau)B_0(\rho_0)) + {\rm Tr}(A {\cal G}(-\tau)B_0(\rho_0))  \nonumber \\
 & = & {\rm Tr}(A {\cal G}(|\tau|)B_0(\rho_0)) \nonumber \\
 & = & 2{\rm Tr}(A {\cal G}(|\tau|){\cal A}\rho_0) -  2{\rm Tr}({\cal A} \rho_0) {\rm Tr}( A{\cal G}(|\tau|)\rho_0) \nonumber \\
 & = &  2{\rm Tr}(A {\cal G}(|\tau|){\cal A}\rho_0) -  2{\rm Tr}({\cal G}(|\tau|){\cal A} \rho_0) {\rm Tr}( A\rho_0) \nonumber \\
 & = & 2{\rm Tr}[(A  - {\rm Tr}(A\rho_0)){\cal G}(|\tau|){\cal A}\rho_0]. \label{QMautocorrelation}
 \end{eqnarray}
 The identity $ {\rm Tr}({\cal G}(|\tau|){\cal A} \rho_0) = {\rm Tr}({\cal A} \rho_0)$ used after line 3 holds since ${\cal G}(|\tau|)$ strictly conserves the trace. Moreover, ${\cal G}(|\tau|)\rho_0 = \rho_0$ holds since $\rho_0$ is the steady state.
 After Fourier-transformation of $G(\tau)$ we find
 \begin{equation}
   S_z^{(2)}(\omega) =  \beta^2 S_{\rm sn} + \beta^4 S_{\rm q} + O(\beta^6) \label{O6S}
\end{equation}
with
\begin{equation}
 S_{\rm sn} = \frac{1}{4}
\end{equation}
and
\begin{equation}
S_{\rm q}(\omega)  = \left({\rm Tr}\left[ (A - {\rm Tr}[A \rho_0])({\cal K}(\omega) + {\cal K}(-\omega) )\rho_0\right]\right) \nonumber
\end{equation}
being the desired noise power spectrum of the quantum system where we defined the superoperator ${\cal K}(\omega)\rho = {\cal G}(\omega)
(A \rho + \rho A)/2$.
Since in the time domain $x(t) := {\rm Tr}\left[ (A - {\rm Tr}[A \rho_0]){\cal K}(t)\rho_0\right]$ is always real valued implying  $x^*(\omega) = x(-\omega)$, $S_{\rm q}(\omega)$ can more compactly be written as
\begin{equation}
S_{\rm q}(\omega)  = \left({\rm Tr}\left[ (A - {\rm Tr}[A \rho_0]){\cal K}(\omega) \rho_0\right] + {\rm c.c.}\right).
\label{spectrumQM}
\end{equation}
The leading order contribution $S_{\rm sn}$ to $S_z^{(2)}(\omega)$ arises from $z_1$ only and can be interpreted as shot noise with a flat spectrum, so-called white noise. This noise contribution is always present even in the absence of the quantum system ($A = 0$).  The fourth order contribution $S_{\rm q}(\omega)$ arises from the correlation of $z_1(\omega)$ and $z_3(\omega)$ and can be interpreted as the leading order noise contribution of the quantum system to $S_z^{\rm (2)}$. 
The contribution of $S_{\rm sn}$ to $S(\omega)$ scales with the laser intensity $\beta^2$,
 while $S_{\rm q}$ grows quadratically with the laser intensity in agreement with observation \cite{mullerPHYSICAE2010}.
Eq.~(\ref{spectrumQM}) is the desired noise spectrum of a general quantum system whose dynamics is given by the very general Lindblad
 master equation $\dot{\rho} = {\cal L}\rho$. The expression allows for the treatment of damping via the Liouvillian ${\cal L}$
 and therefore for a fully quantum mechanical treatment of e.g. spin relaxation. 
 A corresponding derivation of $S_z^{(2)}(\omega)$ for the special case of a two level system was previously given by Korotkov \cite{korotkovPRB2001}. 
 
 An earlier alternative derivation of $S_z^{(2)}(\omega)$ for general systems was given by Barchielli {\it et al.} via a generating function approach and the SME \cite{barchielliBOOK2009,barchielliQM2013}. However, no higher moments or cumulants were explicitly calculated. Their expression for $S_z^{(2)}(\omega)$ implied an absence of the $O(\beta^6)$ contribution. They state on page 118 of their book that the characteristic operator [their Eq.~(5.16)] depends only linearly on $\rho_0$ which implies that all moments of $z(t)$ which follow from that operator are also linear in $\rho_0$. This remarkable result was obtained from a fundamental mathematical treatment of the non-linear SME and an equivalent formulation of a linear SME that describes the evolution of an unnormalized  density matrix. The correct treatment of probability measures eventually led them to their important result. 
Considering that our expressions for $z_j(t)$ depend non-linearly on $\rho_0$, a simple linear dependence for moments of $z(t)$ is not trivially expected. In Section \ref{sec:Multitime} we, however, derive expressions for multi-time moments of $z(t)$ and find that the moments for $n$ different times depend always linearly on $\rho_0$ and $\beta^n$ without any higher order corrections. Consequently, the higher order contributions $O(\beta^6)$ in Eqs.~(\ref{O6G}) and (\ref{O6S}) are identical to zero in agreement with Barchielli. The expression for $S_{\rm q}$, Eq.~(\ref{spectrumQM}), is therefore correct for arbitrary measurement strength. Note that ${\cal K}$ does change with increasing $\beta$ since ${\cal L}$ includes a damping term $-\beta^2[A,[A,\rho]]/2$.
 The relation of $S_{\rm q}$ to traditional
 expressions for the quantum noise based on the Landau approach is established in the next section. 
 
\section{Comparison of $S_{\rm q}(\omega)$ with Landau's approach}
\label{sec:LandauApproach}
In this section we compare the formula for $S_{\rm q}(\omega)$ derived above with Landau's formula Eq.~(\ref{LandauFormula1}). In a recent review, Clerk {\it et al.} restate Landau's formula as   \cite{clerkRMP2010}
\begin{equation}
   S^{\rm (L)}(\omega) = \int_{-\infty}^{\infty} G^{\rm (L)}(\tau) e^{i \omega \tau }\,{\rm d}\tau \label{spectrumClerk}
\end{equation}
with
\begin{equation}
  G^{\rm (L)}(\tau) = \frac{1}{2}{\rm Tr}([A(\tau) A + A A(\tau)]\rho_0). \label{LandauFormula}
\end{equation}
calling it a ‘symmetrized quantum noise spectral density’.
We note that Eq.~(\ref{spectrumClerk}) does  allow for the treatment of damping only after $\rho_0$ has been replaced by a density matrix $\chi_0$ that represents both system and a bath that is coupled to the system. Simply keeping $\rho_0$ in the formula and working with a damped operator $A(t)$ is not correct:
While $A(\tau)$ may relax to some equilibrium value 
 for $\tau \rightarrow \infty$, it can diverge for $\tau \rightarrow -\infty$. Even if $A(\tau)$ does not diverge, 
 temporal symmetry is broken by damping and $ G^{\rm (L)}(\tau) \neq   G^{\rm (L)}(-\tau)$ follows. 
Temporal asymmetry  leads to a complex-valued $S^{\rm (L)}(\omega)$ while any second order noise spectrum
 is required to be real-valued. 
 Interestingly, the authors of Refs. \onlinecite{clerkRMP2010,sinitsynRPP2016} used Eq.~(\ref{spectrumClerk})
and neglected damping of $A$ for calculating $S^{\rm (L)}(\omega)$. They later introduced damping by 
  hand replacing delta-functions in the spectrum by Lorentz-profiles. 
 Next, we show that an expression similar to $G^{\rm (L)}(\tau)$ that allows for treatment of damping for the reduced density matrix $\rho_0$ follows from z-theory and alternatively  from the quantum regression theorem (see next Section). 

Starting from the third line of Eq.~(\ref{QMautocorrelation}) we find
\begin{eqnarray}
  G_{\rm q}(\tau)  & =  & {\rm Tr}(A {\cal G}(|\tau|)( A\rho_0 + \rho_0 A))/2  \nonumber \\
  & & -  {\rm Tr}(A \rho_0) {\rm Tr}( A{\cal G}(|\tau|)\rho_0) \nonumber \\ 
   & = &  \frac{1}{2}\langle A(|\tau|) A(0) + A(0) A(|\tau|) \rangle -  \langle A(|\tau|)\rangle \langle A(0) \rangle \nonumber\\
   & = &  \frac{1}{2}\langle A(|\tau|) A(0) + A(0) A(|\tau|) \rangle -  \langle A(0) \rangle^2  \label{QMautocorrelation2}
\end{eqnarray}
where we switched from the Schr\"odinger  into the Heisenberg picture. The new expression obviously fulfills temporal symmetry $G_{\rm q}(\tau) = G_{\rm q}(-\tau)$ even in the presence of a damped operator $A(\tau)$. The last line follows since $\langle A(|\tau|)\rangle = {\rm Tr}(A(|\tau|) \rho_0)$ does not depend on time for $\rho_0$ in equilibrium. The equation of motion for $A(t)$ in the Heisenberg picture can according to Lindblad be deduced from the original master equation for $\rho$ [Eq.~(\ref{SME_zero})]   as long as it is of the very general Lindblad type \cite{lindbladSPRINGER1976}. Consequently, the treatment of damping is possible in both the Schr\"odinger and the Heisenberg picture. 
The Landau form $G^{\rm (L)}(\tau)$ (used in \onlinecite{landauBOOKstat} for proving the fluctuation dissipation theorem) is easily recovered from
the first term of Eq.~(\ref{QMautocorrelation}) if damping is neglected and $\langle A \rangle = 0$ (see Appendix 
\ref{app:Equivalence}). The  $-  \langle A(0) \rangle^2 $ contribution to $G_{\rm q}(\tau)$ guarantees that our $S_{\rm q}(\omega)$ has no delta-like contribution at $\omega = 0$ even if the usually assumed relation $\langle A \rangle = 0$ does not hold. 

To conclude the section, we want to emphasize that the fluctuation dissipation theorem (FDT) formulated with the traditional definition of the noise spectrum [Eq.~(\ref{spectrumClerk})]
\begin{equation}
 {\rm Im} \,\alpha(\omega) = \frac{1}{\hbar} S^{\rm (L)}(\omega)  \frac{1- e^{-\hbar\omega/k_{\rm B}T}}{1+e^{-\hbar \omega/k_{\rm B}T}}.
\end{equation}
with the complex susceptibility $\alpha(\omega)$ holds in the same form
\begin{equation}
 {\rm Im} \,\alpha(\omega) = \frac{1}{\hbar} S_{\rm q}(\omega)  \frac{1- e^{-\hbar\omega/k_{\rm B}T}}{1+e^{-\hbar \omega/k_{\rm B}T}}.
\end{equation}
 also for our cumulant based $S_{\rm q}(\omega)$ [Eq.~(\ref{spectrumQM})]. 
 The difference $S^{\rm (L)}(\omega) - S_{\rm q}(\omega) = 2 \pi \delta(\omega)  \langle A(0) \rangle^2$ has only spectral weight at $\omega = 0$. The spectral weight at $\omega = 0$ is not important in the fluctuation dissipation theorem as the factor $((1 - \exp (-\hbar \omega / k_{\rm B} T)) = \frac{\hbar \omega}{k T} + O(\omega^2)$ completely suppresses any effect on the above equation since $\int_{-\varepsilon}^\varepsilon \delta(\omega) \omega \,{\rm d}\omega = 0$. A quantum mechanical derivation of the FDT using our notation is given for completeness in Appendix \ref{app:FDT}. 
 
\section{Quantum regression theorem}
\label{sec:QRT}
We found above as a first result of our theory that the quantum mechanical noise spectrum can be expressed as the Fourier transformation of $G(\tau)$ which includes damping for the reduced density matrix $\rho_0$. We will validate our result in this section by deriving basically the same expression with the help of the quantum regression theorem (QRT) starting from $G^{\rm (L)}(\tau)$ \cite{laxPR1963}.
The effects of damping on $A$ can be treated within Landau's approach if instead of the system density matrix $\rho_0$ the density matrix $\chi_0$ of the system (S) plus a reservoir (R) is regarded which causes relaxation of the system via coupling.
The new $G^{\rm (L)}$ consequently reads
\begin{equation}
  G^{\rm (L)}(t'-t)  =  \frac{1}{2}{\rm Tr}_{\rm S+R}[\chi_0 (A(t) A(t') + A(t') A(t))] \label{newLandau}
\end{equation}
where $\chi_0$ and $A$ refer now to the full system S + R.
The quantum regression theorem \cite{laxPR1963} provides a way to simplify the calculation of 
any two-time correlator
\begin{equation}
 \langle O_1(t) O_2(t') \rangle_{\rm S + R} := {\rm Tr}_{\rm S + R}(\chi_0 O_1(t) O_2(t'))
\end{equation}
where $O_j$ are operators that belong to the system part S. 
The quantum regression theorem states that the expectation value can be evaluated in Markov approximation by considering the master equation for the reduced density matrix $\dot{\rho}  = {\cal L}\rho$ where damping and coherent evolution is described by the superoperator ${\cal L}$.
Carmichael gives the following approximate expressions \cite{carmichaelBOOK1993}
\begin{eqnarray}
   \langle O_1(t) O_2(t') \rangle_{\rm S+R} & = & {\rm Tr}_{\mathrm S}[ O_1 e^{{\cal L}(t'-t)}( O_2 \rho(t))] \,\, 
                    \text{for $\,t'> t$} \nonumber \\
                       \langle O_1(t) O_2(t') \rangle_{\rm S+R}  & = & {\rm Tr}_{\mathrm S}[O_1 e^{{\cal L}(t-t')}( \rho(t) O_2 )] \,\, 
                    \text{for $\,t> t'$}  \nonumber \\
                    & & \label{QRT}
\end{eqnarray} 
where the RHSs are formulated in the Schr\"odinger picture. Assuming thermal equilibrium, $\rho(t) = \rho_0 = {\rm const}$, we rewrite the RHSs in the Heisenberg picture for the system S including damping and obtain
\begin{eqnarray}
   \langle O_1(t) O_2(t') \rangle_{\rm S+ R} & = & \langle O_1(t'-t) O_2  \rangle_{\rm S} \,\, 
                    \text{for $\,t'> t$} \nonumber \\
                       \langle O_1(t) O_2(t') \rangle_{\rm S+R} & = & \langle O_2 O_1(t-t')\rangle_{\rm S} \,\,
                    \text{for $\,t> t'$}. \nonumber \\
                    & & 
\end{eqnarray} 
Applying the equations above to Landau's formula [Eq.~(\ref{newLandau})], yields the remarkably simple result
\begin{eqnarray}
  G^{\rm (L)}(t'-t) & = & \frac{1}{2}\langle A(t) A(t') + A(t') A(t) \rangle_{S+R} \nonumber \\
  &  & \hspace{-1.5cm}= \frac{1}{2}\langle A(0) A(|t'-t|) + A(|t'-t|) A(0) \rangle_{\rm S} \label{AutoRegression}
\end{eqnarray}
which for $\langle A(\tau) \rangle = 0 $ is identical to our expression for $G_{\rm q}(\tau)$ obtained 
from z-theory [Eq.~(\ref{QMautocorrelation2})].
While the application of the quantum regression theorem to Landau's formula for arriving at Eq.~(\ref{AutoRegression}) seems very natural, we are not aware of a any comparable derivation in the literature. 
Even the recent review on quantum noise by Clerk {\it et al.} states only the original version of the Landau formula Eq.~(\ref{spectrumClerk}) and does not mention the QRT.
Usually the application of the QRT is done in the Schr\"odinger picture as given by Eq.~(\ref{QRT}) (compare e.g. the use of the QRT by Carmichael \cite{carmichaelBOOK1993}) and consequently does not directly yield equations like Eq.~(\ref{AutoRegression}) formulated in the Heisenberg picture. 
\section{Measurement induced damping in a spin noise experiment and the quantum Zeno-effect}
\label{sec:Zeno}
In the following we investigate the appearance of the quantum Zeno-effect in a spin noise experiment. The measurement induced damping term $-\frac{\beta^2}{2}[A,[A,\rho]]$ will turn out to completely dominate the system dynamics for an increasing measurement strength $\beta$. We consider a single electron spin in an external magnetic field.
The stochastic master equation
\begin{eqnarray}
 {\rm d}{\rho}   &=  &\frac{i}{\hbar}[\rho,H]\, {\rm d}t + {\cal D}\rho \,{\rm d}t  - \frac{\beta^2}{2}[A,[A,\rho]] \,{\rm d}t  \nonumber \\
           & & + \beta [A\rho  + \rho A - 2 \rho {\rm Tr}(\rho A)]\,{\rm d}W 
\end{eqnarray}
will be rewritten into an equivalent quasi-classical equation that allows for a simple interpretation.  
The coherent contribution to the electron spin dynamics in a magnetic field is given by the Hamiltonian 
\begin{equation}
 H = \frac{1}{2}\sum_{ j=x,y,z} \hbar \omega_j \sigma_j,
 \end{equation} 
where $\sigma_j$ are the Pauli spin matrices and $\hbar \omega_j$ are the orientation dependent spin splitting energies.
The spin relaxation due to coupling with the environment 
\begin{equation}
 {\cal D} \rho = - \gamma (\rho - {\rm Tr}(\rho)\mathbbm{1}/2)
\end{equation}
is assumed to be isotropic \cite{bussPRB2016}.
The term that describes measurement induced damping simplifies for $A = \sigma_z$ to
\begin{equation}
- \frac{\beta^2}{2}[A,[A,\rho]] =  -\beta^2(\rho - \sigma_z \rho \sigma_z).
\end{equation}

The time dependent density matrix $\rho(t)$ can for the case of a single electron spin be decomposed into four contributions via 
\begin{equation}
 \rho(t) = \mathbbm{1}/2 +  \frac{1}{2}\sum_{j=x,y,z} s_j(t) \sigma_j, \label{eq:decomposed}
\end{equation}
where the scalar quantities $s_j(t)$ have the meaning of spin expectation values, 
$s_j(t) = {\rm Tr}(\rho(t)\sigma_j)$.
Using Eq.~(\ref{eq:decomposed}) the stochastic master equation assumes without any approximations the form
\begin{eqnarray}
 {\rm d}{\vec{s}}   = \left[ \vec{\omega} \times \vec{s} - \gamma \vec{s} - 2 \beta^2 \left( \begin{array}{ccc} 1 & 0 & 0 \\ 0 & 1 & 0 \\ 0 & 0 & 0 \end{array}\right) \vec{s} \right] \, {\rm d}t  \nonumber \\
  + 2\beta  \left[ \left( \begin{array}{c} 0 \\ 0  \\ 1 \end{array}\right) -  s_z \vec{s} \right] \,{\rm d}W, \label{langevinlike1}
\end{eqnarray}
while the detector output becomes
\begin{equation}
  z(t) = \beta^2 s_z(t) + \frac{1}{2} \beta \Gamma(t). \label{langevinlike2}
\end{equation}
The measurement induced damping of the system [third term of RHS of Eq.~(\ref{langevinlike1})] can now easily be interpreted as an anisotropic spin dephasing of the $x$ and $y$ spin component towards zero with a dephasing rate $\gamma^{\rm M}_x = \gamma^{\rm M}_y = 2 \beta^2$ \cite{mullerPRL2008,poltavtsevPRB2014,yangRPP2017}.  
It is also obvious that for an increasing measurement strength $\beta$ and $\gamma^{\rm M}_x > |\vec{\omega}|$ the dynamics corresponds to an overdamped oscillation where spin precession is completely suppressed. This behavior is known as the quantum Zeno-effect where the coherent evolution of a quantum system is suppressed by strong measurements \cite{misraJMP1977}.
 \begin{figure}
   \centering
 \includegraphics[width=8cm]{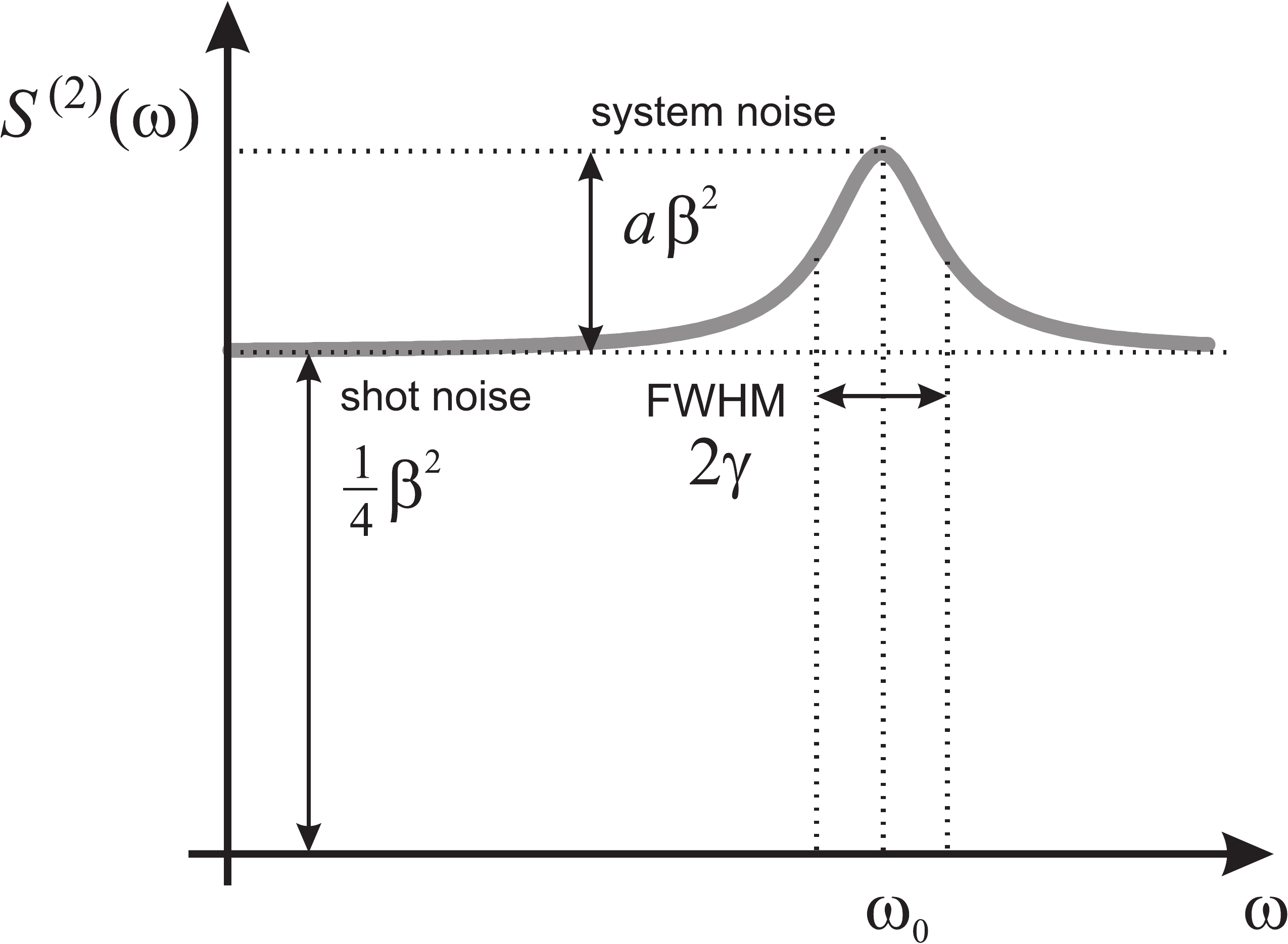} 
   \caption{The quantities $a$ and $\gamma$ can be determined from a measured noise spectrum and allow for an estimate of the measurements strength $\beta$ via $\beta^2 = \gamma a$. }
    \label{spinNoiseSpectrum}
    \end{figure}

Next, we establish a useful connection between the measurement strength $\beta$, the area $I_{\rm q}$ under the quantum noise spectrum $S_{\rm q}$, and the shot noise background $S_{\rm sn}$ (see Figure \ref{spinNoiseSpectrum}). This will allow us to deduce the measurement induced spin dephasing rates $\gamma^{\rm M}_x = \gamma^{\rm M}_y = 2 \beta^2$ from the measured spin noise spectrum $S(\omega) =  \beta^2/4 + \beta^4 S_{\rm q}(\omega)$. 
We use the following property of the integrated quantum mechanical noise [see e.g. Eq. (9.17) in Ref. \cite{barchielliBOOK2009}]
\begin{eqnarray}
 I_{\rm q} & = & \beta^4 \int_{-\infty}^{\infty} S_{\rm q}(\omega) {\rm d}\omega \nonumber \\
  & = & 2 \pi \beta^4 G(0) \nonumber \\
  & = &  2 \pi \beta^4({\rm Tr}( \rho_0 A^2)  -  ({\rm Tr}( \rho_0 A))^2) \nonumber \\
  & = & 2 \pi \beta^4  \label{sumRule}
\end{eqnarray}  
where the second line follows from Eq.~(\ref{QMautocorrelation}) and the last line follows for a spin noise experiment ($A = \sigma_z$) at high temperature where $\rho_0 = \mathbbm{1}/2$. Similar relations can be found also for other measurement operators $A$.
The area $I_{\rm L}$ under a single Lorentz-shaped peak of a spin resonance damped with damping constant $\gamma$ and peak height $a \beta^2 $ is
\begin{equation}
 I_{\rm L} =  a \beta^2 \int_{-\infty}^{\infty} \frac{\gamma^2}{(\omega-\omega_0)^2 + \gamma^2}{\rm d}\omega= a\beta^2 \pi  \gamma,
\end{equation} 
where $a$ is dimensionless.

If a single peak is observed in a noise experiment the relation
\begin{eqnarray}
I_{\rm L} & = & \frac{1}{2} I_{\rm q}
\end{eqnarray}
must hold. The factor $1/2$ appears because $ I_{\rm q}$ represents the area under two noise peaks - one at $\omega_0$ and another at $-\omega_0$.  Eventually we find a relation that yields $\beta^2$ for known $\gamma$ and $a$ 
\begin{eqnarray} 
a \beta^2 \pi \gamma & = & \pi \beta^4  \nonumber \\
 \beta^2 & = &  \gamma a
\end{eqnarray}
that holds for the second order noise spectrum of an arbitrary quantum system.
In the special case of a spin noise experiment, the measurement induced spin dephasing rate
\begin{equation}
 \gamma^{\rm M}_x = 2 \beta^2 = 2 a \gamma. \label{eq:sum_rule}
\end{equation}  
follows from Eq.~(\ref{langevinlike1}). Consequently, $\gamma^{\rm M}_x$  is given by $a$ and $\gamma$ which both can be determined from a spin noise spectrum that was measure with laser power $P_{\rm L}$  (see Figure \ref{spinNoiseSpectrum}). 
The knowledge of $\gamma_x^{\rm M}$ can then be used to estimate the required laser power $P^{\rm Zeno}_{\rm L}$ to reach the Zeno-regime. Since $\gamma^{\rm M, Zeno}_x \approx \omega_0$ is required for the Zeno-regime, we find
$P^{\rm Zeno}_{\rm L} = P_{\rm L} \omega_0 /\gamma_x^{\rm M}$.
In general, many spins $N$ may be measured simultaneously if both the laser spot area and the laser power are increased by a factor of $N$. In that case $\beta$ is independent of the spot area. The measured spectrum will show an $N$-fold increase for both the shot-noise background level and the area under the spin noise peak which leaves the parameter $a$ unchanged compared to the single spin measurement.  
In practice, a fraction $p$ of the laser beam may miss a spin and cause an increase in the shot noise background that is uncorrelated with the spin measurement. In that case the relation Eq. (\ref{eq:sum_rule}) no longer holds exactly and should be replaced by $\beta^2 =  p \gamma a$. A distribution of spins in a two-dimensional plane with a spacing of roughly a wavelength would correspond to about $p = 1$.
 
 The Zeno-effect may have already been observed in a semiconductor spin noise experiment \cite{poltavtsevPRB2014}, however, left unnoticed by the authors. Their Figure 3 shows a clear transition from a spin precession peak at a finite frequency to a broad peak at zero frequency for increasing probe laser intensities as expected for the Zeno effect. Their semiconductor single quantum well was placed into an optical microcavity which led to a strong field enhancement of the probe laser beam resulting in a relatively low probe power of 3~mW where the effect appeared. Values between about 1.25 and 3 for our parameter $a$, $\gamma \approx  \omega_0/3$, and $p \approx 1$ can be estimated from their data which supports our interpretation of their experiment as a manifestation of the quantum Zeno effect.
 \section{Stochastic Master Equation versus Langevin-approaches}
\label{sec:Langevin}
In 1908, Paul Langevin initiated a new theoretical treatment of Brownian motion in terms of a differential equation with a stochastic driving term \cite{langevinCR1908,uhlenbeckPR1930}. So-called Langevin-approaches have since then become very popular for the stochastic treatment of diffusion and damping in many branches of physics.
Glazov and Ivchenko calculate spin noise spectra of a single spin from the following Langevin-equation
\cite{glazovPRB2012}
\begin{equation}
 \frac{\partial \delta \vec{s}(t)}{\partial t} +  \frac{\delta \vec{s}(t)}{\tau_s} +  \delta \vec{s}(t) \times \omega_{B} = \vec{\xi}(t), \label{GlazovLangevin}
\end{equation}
where $\vec{\xi}(t)$ is a stochastic Gaussian driving term.
The detector output in their theory
\begin{equation}
 \tilde{z}(t) = ( \delta \vec{s}(t) )_z \label{GlazovLangevin2}
\end{equation}
was assumed to be identical with the z-component of the fluctuating spin vector.
A comparison with our single spin theory [Eqs.~(\ref{langevinlike1}) and (\ref{langevinlike2})] exhibits similarities and a few striking differences. Glazov's equations are linear in $\delta \vec{s}$ while our equations are non-linear in $\vec{s}$. The stochastic Gaussian driving term $\vec{\xi}$ can therefore cause only a Gaussian response of $\delta \vec{s}$. Consequently, all higher order spectra of $( \delta \vec{s}(t) )_z$ vanish. In contrast, our theory exhibits a non-linear coupling of the stochastic quantity $\Gamma(t)$ to $\vec{s}$ which in general will lead to non-vanishing higher order spectra. The stochastic driving term of the Langevin theory $\vec{\xi}(t)$ scales according to $\langle \xi_j(t’)\xi_j(t) \rangle = \tau_0 \delta(t-t’) /2$ where the prefactor must be chosen in a way to drive spin fluctuations whose value of $\delta \vec{s}^2$ is consistent with the $3/4$ value in thermal equilibrium. In contrast, the driving term of the stochastic master equation is proportional to the measurement strength $\beta$. Particularly in the limit $\beta \rightarrow 0$ the desired noise spectrum $S_q(\omega)$ for a vanishing perturbation by the probe beam is obtained. 
The correct value for $\vec{\sigma}^2$ (which corresponds to $\delta \vec{s}^2$) is in the case of the SME is built into the steady state density matrix $\rho_0$ which results from $H$ and coupling to the environment via ${\cal D}$ (case of weak measurement, where $\beta \rightarrow 0$).
Clearly, the stochastic driving term of the Langevin approach and of the SME have completely different meanings and it cannot be expected that the Langevin approach might appear after some approximation from the SME.

Nevertheless, the traditional Langevin approach of Eq.~(\ref{GlazovLangevin}) led Glazov to results consistent with the results of a fully quantum mechanical treatment by Braun \cite{braunPRB2007} and also consistent with our result Eq. (\ref{eq:SimpleNoise}). A linear Langevin-theory will, nevertheless, always result in vanishing higher order spectra.

\section{Examples of spin noise spectra}
\label{sec:Examples}
Here we calculate the spin noise spectrum  $S_{\rm q}(\omega)$ of the $z$ component of a singe electron spin that is precessing in an external magnetic field oriented in $x$-direction and subject to relaxation.
We first calculate 
\begin{equation}
  x(t) = {\cal G}(t)(A\rho_0 + \rho_0 A)
\end{equation}
that we will need to evaluate ${\cal K}(\omega)\rho_0$ of Eq.~(\ref{spectrumQM}).
For $t \ge 0$ we find for $A = \sigma_z$ and the high temperature limit $\rho_0 = \mathbbm{1}/2$
\begin{equation}
  x(\tau) = e^{{\cal L}\tau}\sigma_z.
\end{equation}
The propagation of $\sigma_z$ can be treated with the help of Eqs.~(\ref{eq:decomposed}) and (\ref{langevinlike1}) by noting that $\sigma_z$ corresponds to $\vec{s}(0)^{\rm T} = (0,0,2)$ that propagates in time
according to
\begin{equation}
  \dot{\vec{s}}   =  \vec{\omega} \times \vec{s} - \gamma \vec{s},
\end{equation} 
where we neglected the measurement induced spin dephasing.
The time dependent solution
\begin{equation}
  \vec{s}(\tau) = \left(\begin{array}{c} 0 \\ 0  \\ 2 \end{array}\right) \cos (\omega_x \tau) e^{-\gamma t}
- \left(\begin{array}{c} 0 \\ 2  \\ 0 \end{array}\right) \sin(\omega_x \tau) e^{-\gamma t}
\end{equation}
is equivalent to 
\begin{equation}
 x(\tau) = \sigma_z \cos (\omega_x \tau)e^{-\gamma \tau} - \sigma_y \sin(\omega_x \tau) e^{-\gamma \tau}.
\end{equation}
This leads us to
\begin{eqnarray}
 S_{\rm q}(\omega) & = & \frac{1}{2}({\rm Tr}(\sigma_z x(\omega)) + {\rm c.c.}) \nonumber \\
 & = & \frac{1}{2}\left( \int_0^\infty 2 \cos(\omega_x \tau) e^{-\gamma \tau} e^{i\omega \tau} {\rm d}\tau+ {\rm c.c.}\right) 
 \nonumber\\
 & = & \frac{1}{2}\left(  \frac{-1}{i(\omega + \omega_x) - \gamma} + \frac{-1}{i(\omega - \omega_x)-\gamma}\right) \nonumber \\
 & & + \,\,{\rm c.c.} \nonumber \\
 & = & \frac{\gamma}{(\omega + \omega_x)^2 + \gamma^2} +\frac{\gamma}{(\omega - \omega_x)^2 + \gamma^2} \label{eq:SimpleNoise}
\end{eqnarray}
which is equivalent to Braun's equation (A4) \cite{braunPRB2007}. The area under the two peaks is $2 \pi$ as 
expected from Eq.~(\ref{sumRule}).

 \begin{figure}
   \centering
 \includegraphics[width=8cm]{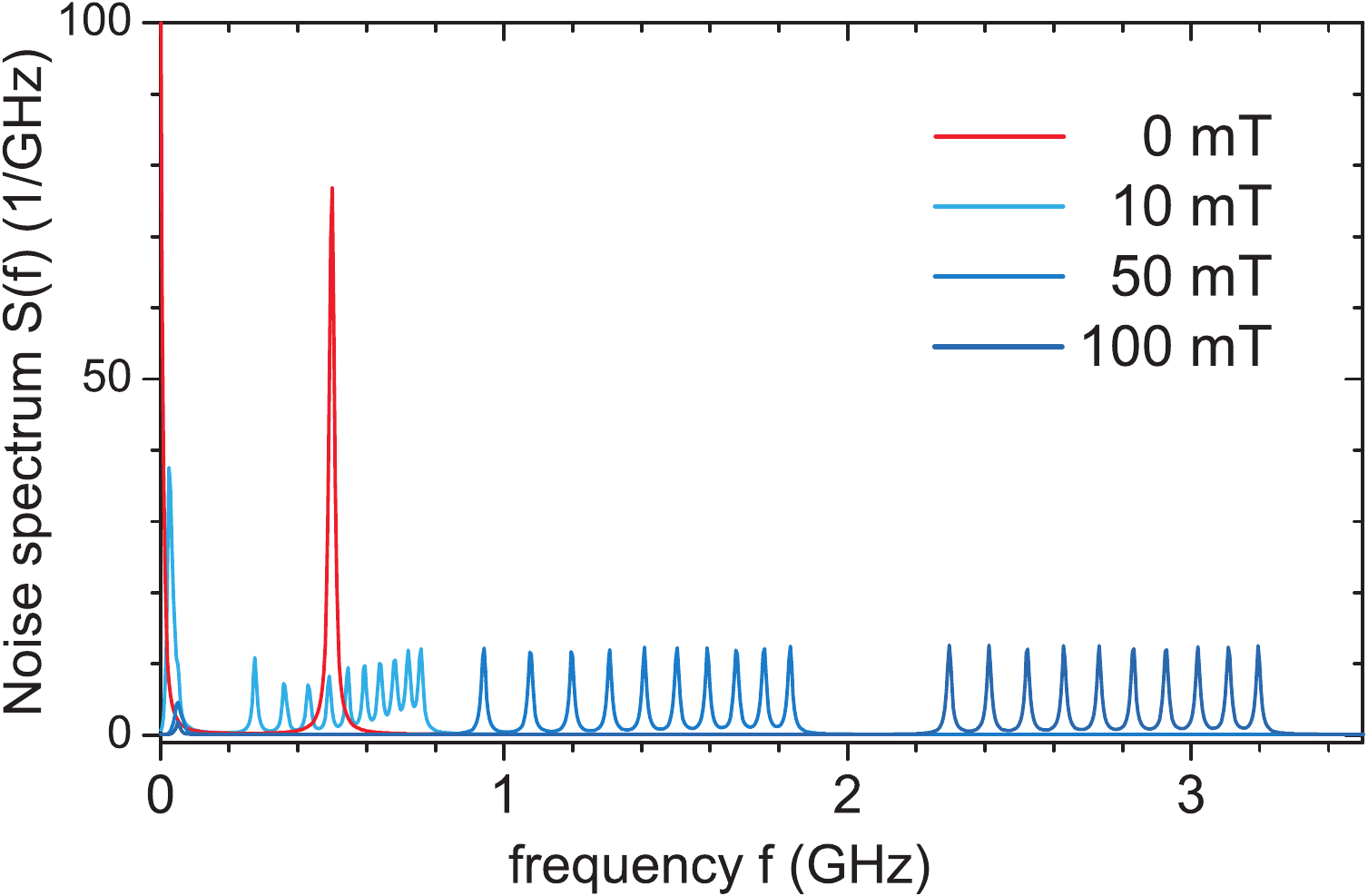} 
   \caption{Calculated spin noise spectra of the coupled spin system in ZnO:In for high spin temperatures and increasing magnetic fields. }
    \label{spinNoiseSpectrumZnO}
\end{figure}
Next, we numerically calculate spin noise spectra of a coupled spin spin system. Higher order spectra of the same system will given in Section \ref{sec:S4spinpair}.
The indium donor in the semiconductor ZnO exhibits a strongly coupled spin pair consisting of the $I = 9/2$ indium nuclear spin and
the $s = 1/2$ electron spin of the localized electron donor \cite{bussPRB2016}.  
The dynamics of the interacting electron spin $\vec{s}$ and nuclear spin $\vec{I}$ in a 
magnetic field $\vec{B}$ is given by the Hamiltonian\cite{blockPRB82}
\begin{equation}
  H = \beta g^{\rm (e)} \vec{B}\cdot \vec{s} + A \vec{I}\cdot \vec{s} + P_{\parallel} I_z^2 - \beta g^{\rm (n)} \vec{B} \cdot \vec{I},
\end{equation}
where the first and last term describe the electronic and nuclear spin precession in the external magnetic field $\vec{B}$. The hyperfine 
coupling is described by the second term. The third term is due to a weak electric quadrupole crystal field splitting that in 
principle can be exploited for spin squeezing.\cite{kitagawaPRA1993}
The following parameters are known from ENDOR experiments (see Ref. \onlinecite{blockPRB82}): $\beta g^{\rm (e)}/\hbar = 0.172\times 10^{12}$~{\rm rad~s}$^{-1}$~T$^{-1}$ ($g^{\rm (e)} = 1.96$), $\beta g^{\rm (n)}  / \hbar = 9.329 \times 10^6$~{rad~s}$^{-1}$~T$^{-1}$, 
$A/h = 100.2$~MHz, and $P_\parallel/h = 1.27$~MHz.   The electron-nuclear system is fully described by a density matrix $\rho$ in the 20-dimensional combined Hilbertspace $H_n \otimes H_e$ for nuclear spin and electron spin. The master equation
\begin{eqnarray}
    \dot{\rho} & = & {\cal L} \rho \nonumber \\ 
 & = & \frac{i}{\hbar} \left[  \rho, H \right ]  + \Gamma^{\rm (e)}_{\rm relax} + \Gamma^{\rm (n)}_{\rm relax} 
  \label{masterequation}
\end{eqnarray}  
describes both the coherent propagation (first term) and dissipative coupling to the environment (last two terms). 
The relaxation of the electron spin towards its equilibrium orientation 
$\rho^{\rm (e)}_{\rm final} \propto \exp(-\beta g^{\rm (e)} \vec{B}\vec{s}/k_{\rm B} T)$ is in the most simple form of 
isotropic relaxation given by \cite{bussPRB2016}
\begin{equation}
\Gamma^{\rm (e)}_{\rm relax} =  - \gamma_{\rm relax} \left[  \rho - ({\rm Tr}_e \rho)\otimes \rho^{\rm (e)}_{\rm final} \right],
\end{equation}
with the spin relaxation rate $\gamma_{\mathrm{relax}}$.
The temperature dependence of $\rho^{\rm (e)}_{\rm final}$ leads to a temperature dependence of $\rho_0$ which allows for the calculation of noise spectra for different system temperatures. 
The symbol ${\rm Tr}_e$ denotes the partial trace over the density matrix with respect to the electronic subsystem. 
The nuclear state ${\rm Tr}_e \rho$ may be interpreted as a spin state that has lost all entanglement with the electronic state.
Similarly, nuclear spin relaxation is modelled via 
\begin{equation}
\Gamma^{\rm (n)}_{\rm relax} =  - \gamma_{\rm relax} \left[  \rho - \rho^{\rm (n)}_{\rm final}\otimes ({\rm Tr}_n \rho) \right].
\end{equation}
Fig. \ref{spinNoiseSpectrumZnO} shows spin noise spectra for $\gamma^{\rm (e)}_{\rm relax} = 1/20$~ns$^{-1}$,
$\gamma^{\rm (n)}_{\rm relax} = 1/20$~$\mu$s$^{-1}$, and $\rho_0 \propto \mathbbm{1}$ (high temperature limit) and negligible
measurement induced dephasing 
obtained from Eq.~(\ref{spectrumQM}).
For $B = 100$~mT we find 10 peaks around 3~GHz with a spacing of $0.1$~GHz. These peaks correspond to the 10 nuclear spin levels that split the electron spin resonance due to the hyperfine field. 
For $B = 0$~mT we find a peak at $0.5$~GHz which corresponds to the dynamics of electron spin and nuclear spin in their mutual hyperfine fields with a precession frequency of $(I + 1/2) A$ for an $I = 9/2$ nuclear spin (a short theory of the precession frequency is given in \cite{bussPRB2016}).   
A second peak at zero frequency is explained by an approximately collinear arrangement of the spins along the $\pm z$-direction. Despite a (weak) precession in the hyperfine field the electron spin can roughly keep the orientation along $\pm z$  giving rise to a zero frequency contribution to the spectrum. At a moderate field of 10~mT the low frequency peak shifts to higher frequencies corresponding to a common precession of the strongly coupled spin-spin system around the magnetic field. The high frequency peak splits into ten smaller peaks.
\section{Signal to Noise of $S_z^{(n)}$}
\label{sec:SignalToNoise}
Up to now we considered measurements on a single quantum system. This situation is typical for noise measurements in nano-electronics. In the case of a spin noise experiment, the laser beam often probes many systems that are independent of each other but exhibit basically identical dynamical properties (e.g. electron spins in GaAs localized at silicon donor sites \cite{oestreichPRL2005}). We therefore shortly discuss the situation of $N$ identical systems that are measured with the same detector.
A possible coupling of the systems via the laser is here neglected as typical Faraday angles are usually below $10^{-4}~{\rm rad}$  
which means that the individual systems exchange basically no information on their actual state \cite{oestreichPRL2005}.
In the case of a single system the parameter $\beta$ scales with the laser intensity $I_{\rm L}$ as $\beta \propto I_{\rm L}^{1/2}$ (see Section \ref{sec:SME}).
If the laser spot size area on the sample is increased by a factor of $N$ to probe $N$ systems at the same time, the intensity for a single system is effectively reduced by a factor of $N$. It follows that $\beta$ scales as $\beta \propto N^{-1/2}$.
The new signal $z$ on the detector is then the sum of all individual signals
\begin{equation}
 z(t) = \sum_j z^{(j)}(t).
\end{equation} 
Assuming that all individual systems are independent, it follows from Eq.~(\ref{cumulantsLin}) that the $n$th-order spectrum is given by
\begin{equation}
   S_z^{(n)} = \sum_j    S_{z^{(j)}}^{(n)}. 
\end{equation}
The question now arises if a better signal to noise ratio can be obtained from measuring $N$ systems at the 
same time with laser power $I_{\rm L}$ or a single system with laser power $I_{\rm L}/N$.

In a real experiment the data stream $z(t)$ is divided into $M$ time frames. Each time frame $j$ yields Fourier components $z_j(\omega)$
Usually an estimator $\tilde{S}^{(n)}_{k}$ can be defined from $m$ subsequent frames. A possible (but not the only) estimator for $S_z^{(2)}(\omega)$  is e.g.
\begin{eqnarray}
 \tilde{S}^{(2)}_k  & = & \frac{m}{m-1}\left[  \frac{1}{m} \sum_{k'=0}^{m-1} z^*_{m k+k'}(\omega)    z_{m k+k'}(\omega) \right. \nonumber \\
 & &\hspace{-1cm} - \left. \frac{1}{m^2}\left(  \sum_{k'=0}^{m-1} z^*_{m k+k'}(\omega)\right)  \left(  \sum_{k'=0}^{m-1} z_{m k+k'}(\omega)\right)     \right].
\end{eqnarray}
An estimator for a fourth order spectrum can be found in Ref. \onlinecite{starosielecRSI2010}. For a general theory of estimators see Ref. \onlinecite{fisherPLMS1930}.
The average 
\begin{equation}
  S'^{(n)} =  \frac{m}{M} \sum_{k=1}^{M/m }\tilde{S}^{(n)}_k.
\end{equation}
will then converge to the true $S_z^{(n)}$ for large $M$. We are now interested in the variance $\sigma^2$ of $S'^{(n)}$ as it determines the noise on the measured spectrum $S'^{(n)}$. We assume that the $z_j(\omega)$s are dominated by white photon shot noise $x_j = \sqrt{N} x_0 g_j$ where $g_j$ is Gaussian distributed and $x_0$ regards some scaling factor of the measurement setup. The additional 
scaling factor $\sqrt{N}$ regards the rules under which several equally intense Gaussian noise sources add to a new Gaussian noise source. 
Since $\sigma^2_{\tilde{S}_k} = \langle  \tilde{S}_k^2 \rangle - \langle  \tilde{S}_k \rangle \langle  \tilde{S}_k \rangle \propto x_0^{2n} N^n$, we find 
\begin{equation}
  \sigma^2 \propto x_0^{2n} N^n/M
\end{equation}
while the overall signal $S_z^{(n)}$ scales as $x_0^n N$.
Consequently, the signal to noise ratio of a measured spectrum  $S_z^{(n)}$  (or more precisely $S'^{(n)}$ ) scales for dominant photon shot noise as
\begin{equation}
  S.N. = \frac{S_z^{(n)}}{\sigma} \propto \frac{\sqrt{M}}{N^{n/2 -1}}.
\end{equation}
For $n=2$ there is no advantage in measuring a single instead of $N$ quantum system, whereas for $n = 3$ the gain in signal to noise is a factor of $\sqrt{N}$ and for $n=4$ a factor of $N$. Our $n=4$ result is in agreement with Li's statement that the measurement time (proportional to the number $M$ of time frames) unfavourably increases with $N^2$ if $N$ systems instead of one are measured simultaneously \cite{liPRA2016}.

\section{Multi-time moments of $z(t)$}
\label{sec:Multitime}
In the previous sections we  discussed the second order spectrum $S^{(2)}_z(\omega)$ that was obtained from the explicit expression for $z(t)$ with its leading terms $z_1(t)$ to $z_3(t)$ up to order $\beta^3$. Since $z_1(t)$ to $z_3(t)$ are purely Gaussian, all cumulants $C_n$ and spectra $S^{(n)}$ of $z_1(t) + z_2(t) + z_3(t)$ are zero. Consequently, the calculation of the bispectrum $S_z^{(3)}$ requires at least $z_4(t)$ to be taken into account. However, both the calculation of $z_4(t)$ from the CQNF Eq.~(\ref{eq:CQNF}) and the calculation of a multi-time-cumulant of  $z_1(t) + z_2(t) + z_3(t) + z_4(t)$ turn out to be almost intractable. 
Alternatively, we present in this section a method to directly calculate multi-time moments of $z(t)$ 
\begin{equation}
     M_n = \langle z(t_n)\cdots z(t_1) \rangle
\end{equation}
without the need for
an explicit representation of $z(t)$. Surprisingly compact expressions for multi-time cumulants are then obtained from $M_n$ in Section \ref{sec:multicum}.   
The quantum mechanical expressions for $M_n$ will turn out to be linear in $\rho_0$ and of order $\beta^{2n}$. The expressions are 
valid for any desired coupling strength without the need for higher order corrections. This settles the open question in Section \ref{sec:NoisePower} for possible higher order contributions to $C_2(z(t_2),z(t_1))$  of order $O(\beta^6)$ [Eq.~(\ref{O6S})]. 
The calculations of $M_n$ are directly based on the (non-linear) SME [Eqs.~(\ref{SME}) and (\ref{SME2})] and include external as well as measurement induced damping. 
The derivation is therefore more general than the path-integral based derivation of $M_n$ that was given
by Bednorz {\it et al.} in 2012 \cite{bednorzNJP2012}.
\begin{widetext}
Suppose the quantum system is in the state $\rho_1$ at time $t_1$. 
Since $z(t)$ appears from the solution of the SME, the product $z(t_n)\cdots z(t_1) $ depends on the stochastic quantities ${\rm d}W(t)$ for all times between  $t_1$ and $t_n$. The averaging over all these (infinitly many) quantities is required
to obtain $M_n$. In a first step of calculating $M_n$ we will distinguish between averages taken at times $t_j$ and averages taking in the intervals between those discrete times. We will assume the time order $t_n> t_{n-1} > \cdots > t_1$.  
Consider
\begin{eqnarray}
  M_n & = &  \langle z(t_n)\cdots z(t_1) \rangle_{\Gamma_n T_{n-1} \Gamma_{n-1} \cdots T_1 \Gamma_1} \nonumber
\end{eqnarray}
where we indicate averages taken at time $t_j$ by $\Gamma_j$ and averages take in  the interval $t_{j-1}$ to $t_j$ ($t_j > t_{j-1})$ by $T_{j-1}$.The averaging will successively be performed starting with the latest times.
We find with $z(t_n) = {\rm Tr}(A\rho(t_n)) + \beta \Gamma(t_n)/2)$
\begin{eqnarray}
  M_n & = &  \langle (\beta^2 {\rm Tr}(A\rho(t_n)) + \beta \Gamma(t_n)/2) z(t_{n-1}) \cdots z(t_1) \rangle_{\Gamma_n T_{n-1} \Gamma_{n-1} \cdots T_1 \Gamma_1} \nonumber \\
   & = & \langle \beta^2 {\rm Tr}(A\rho(t_n)) z(t_{n-1})  \cdots z(t_1) \rangle_{
   T_{n-1} \Gamma_{n-1} \cdots T_1 \Gamma_1}
\end{eqnarray}
since $\rho(t_n)$ does not depend on $\Gamma(t_n)$ (Ito-calculus where ${\rm d}\rho = \rho(t + {\rm d}t) - \rho(t)$) and $\langle \Gamma(t_n) \rangle = 0$. 
The averaged density matrix $\langle \rho(t) \rangle$ obeys the master equation
\begin{equation}
 \frac{\partial}{\partial t}\langle \rho(t) \rangle = {\cal L } \langle \rho(t) \rangle 
\end{equation}
in the time interval $T_{n-1}$ as can be seen after averaging the non-linear SME [Eq.~(\ref{SME2})].
The average over the time interval $T_{n-1}$ consequently yields for the first factor in the product
\begin{eqnarray}
  \langle {\rm Tr}(A\rho(t_n)) \rangle_{T_{n-1}} & = & {\rm Tr}(A e^{{\cal L}(t_n - t_{n-1})}\rho(t_{n-1} + {\rm d}t)). 
\end{eqnarray}
This leads to
\begin{eqnarray}
  M_n   & = & \langle \beta^2 {\rm Tr}(A{\cal G}(t_n - t_{n-1})\rho(t_{n-1} + {\rm d}t)) z(t_{n-1})  \cdots z(t_1) \rangle_{
   \Gamma_{n-1} T_{n-2} \cdots T_1 \Gamma_1} \nonumber \\
    & = & \langle \beta^2 {\rm Tr}[A{\cal G}(t_n - t_{n-1})\rho(t_{n-1} + {\rm d}t) z(t_{n-1})] z(t_{n-2})  \cdots z(t_1) \rangle_{
   \Gamma_{n-1} T_{n-2} \cdots T_1 \Gamma_1}
\end{eqnarray}
where in the second line we pulled $z(t_{n-1})$ into the trace.
The evaluation procedure dramatically simplifies by using the formula 
\begin{equation}
 \langle \rho(t_j + {\rm d}t) z(t_{j}) \rangle_{\Gamma_{j}} = \frac{\beta^2}{2}( A\rho(t_{j}) + \rho(t_{j}) A)
 \label{eq:Lemma1}
\end{equation}
which is via Eq.~(\ref{SME2}) obtained from
\begin{eqnarray}
\langle \rho(t_j + {\rm d}t) z(t_{j}) \rangle_{\Gamma_{j}} & = & \langle [
\rho(t_{j}) + \beta (A \rho(t_{j}) + \rho(t_{j}) A) {\rm d}W(t_{j})
 - 2 \beta \rho(t_{j})  {\rm Tr}( A\rho(t_{j} ))\, {\rm d}W(t_{j})
 ] \nonumber \\
 & & \times [\beta^2 {\rm Tr}(\rho(t_{j})A) + \frac{\beta}{2}\Gamma(t_{j})
]\, \rangle_{\Gamma_{j}}.
\end{eqnarray}
Since $\rho(t_{j})$ does not depend on ${\rm d}W(t_j)$, the cross terms  (1,2), (2,1), and (3,1) disappear.
The cross terms (1,1) and (3,2) cancel each other regarding that $\langle \Gamma_j {\rm d}W(t_j)\rangle_{\Gamma_j} = 1$. Only the term (2,2) contributes to the RHS of Eq.~(\ref{eq:Lemma1}).
We find with Eq.~(\ref{eq:Lemma1})
\begin{eqnarray}
  M_n   & = & \langle \beta^4{\rm Tr}(A{\cal G}(t_n - t_{n-1}){\cal A}\rho(t_{n-1})) z(t_{n-2})  \cdots z(t_1) \rangle_{
  T_{n-2} \Gamma_{n-2}\cdots T_1 \Gamma_1} \nonumber 
\end{eqnarray}
and eventually after averaging over the remaining pairs of $T_{n-2}$ and $\Gamma_{n-2}$ using the same procedure as above
\begin{eqnarray}
  M_n   & = &  \beta^{2n} {\rm Tr}(A{\cal G}(t_n - t_{n-1}){\cal A}{\cal G}(t_{n-1} - t_{n-2}){\cal A}\cdots{\cal G}(t_2 - t_{1}){\cal A}\rho(t_{1}))  \label{eq:moments1}
\end{eqnarray}
for $t_n > \cdots > t_1$.
In the rest of the paper we will use Eq.~(\ref{eq:moments1}) in the following form
\begin{equation}
  \langle z(t_n) \cdots z(t_1) \rangle = \beta^{2n} {\rm Tr}(A {\cal G}(t_n - t_{n-1}){\cal A} \cdots {\cal G}( t_{2}-t_1) {\cal A}\rho_0 )
  \label{eq:MomentsK}
\end{equation}
where we use the fact that $\rho(t_1) = \rho_0$ at the time of the first measurement.
For unknown time order the following equation is alway correct
\begin{equation}
  \langle z(t_n) \cdots z(t_1) \rangle = \beta^{2n} \sum_{\text{prm. $t_j$}} {\rm Tr}({\cal A} {\cal G}(t_n - t_{n-1}){\cal A} \cdots {\cal G}( t_{2}-t_1) {\cal A}\rho_0 ),
  \label{eq:MomentsKperm}
\end{equation}
where the expression "prm. $t_j$" below the sum means that all variables $t_j$ have to be permuted. We also replaced $A$ by 
the superoperator ${\cal A}$ which is possible due to the cyclicity of the trace. 

\end{widetext}
The quantum mechanical expression for the multi-time moments $M_n$ of the detector output $z(t)$ derived from the non-linear SME 
are an important intermediate result towards the derivation of higher order cumulants. Eq.~(\ref{eq:MomentsK}) fully characterizes all properties of $z(t)$. It is valid without any restrictions on the measurement strength $\beta$. Calculations for different measurements strength need only to regard that the contribution $-\beta^2[A,[A,\rho]] /2$ to ${\cal L}$ changes with $\beta$ which leads to a $\beta$-dependent propagator ${\cal G}(t)$ and a different steady state density matrix $\rho_0$
[Eq.~(\ref{SME_zero})]. Remarkably, the multi-time moments depend linearly on $\rho_0$ despite the fact that $z(t)$ and the SME are nonlinear in $\rho_0$ and $\rho(t)$, respectively. The non-linearity disappears in the case of the $T_j$-averages because the relevant averaged master equation is strictly linear in $\rho$. In the case of the $\Gamma_j$-averages [see Eq.~(\ref{eq:Lemma1})] the factor $z(t_j)$ - which is non-linear in $\rho_0$ - is replaced by an expression that is linear in $\rho(t_j)$. Barchelli {\it et al.} had stated a linear dependence of all moments on $\rho_0$ already from fundamental arguments about the properties of a moment-generating expressions in the context of the SME (see page 118 in Ref. \onlinecite{barchielliBOOK2009}, where they discuss their theorem 4.14). They, however, derived explicit expressions only for $M_1$ and $M_2$.

Considering the $\beta$-orders of the contributions $z_j(t)$ to $z(t)$ and the $\beta$-order of $M_n$ we find that the moment $M_2$ is only sensitive to 
$z_1(t)$, $z_2(t)$, and $z_3(t)$ but does not depend on $z_4(t)$ or any higher order contribution to $z(t)$. Similarly, one easily finds that generally $M_n$ is only sensitive to $z_j(t)$ for $j \le n+1$. This immediately leads us to the following interesting results for the higher order spectra of continuous quantum measurements: While $S^{(3)}$ requires a non-Gaussian contribution to $z(t)$ for being non-zero, $S^{(2)}$ is completely blind to all non-Gaussian contributions $z_4(t)$, $z_5(t)$ etc.

Bednorz {\it et al.} had found a corresponding formula for $M_n$ \cite{bednorzNJP2012}. 
 Their  derivation is less general than the above via the SME and neglects possible coupling
  to an external bath (i.e. ${\cal D} = 0$) and assumed 
 the limit of weak continuous measurements. They derive an expression $P[a(t)]$ for the probability of finding a measurement trajectory $a(t)$. The expression for $P[a(t)]$ has the form of a path integral which is the basis for defining a generating function [Eq.~(18) in Ref.  \onlinecite{bednorzNJP2012}] for calculating moments 
$\langle a(t_1)a(t_2) \cdots a(t_n)\rangle$ of the measurement trajectory.  They eventually find for $t_n \geq t_{n-1} \geq \cdots \geq t_1$ 
\begin{eqnarray}
\langle a(t_1) \cdots a(t_n) \rangle_{\rm q}  & = & \nonumber \\
& & \hspace{-2.5cm} {\rm Tr}[{\cal A} U(t_n,t_{n-1}) \cdots {\cal A}U(t_2,t_1){\cal A}U(t_1,0)\rho_0]. \label{bednorz}
\end{eqnarray}  
The superoperator $U$ is a propagator for the system $\rho$ \cite{bednorzNJP2012,sinitsynRPP2016}. The index ${\rm q}$ means that only contributions of the system but not of the background noise are considered in the moments. One easily finds that Bednorz's expression is completely equivalent to our Eq.~(\ref{eq:MomentsK}) regarding that $a(t) = z(t)/\beta$, $U(t_2,t_1) {\cal A} = {\cal G}(t_2-t_1){\cal A}$, and $U(t_1,0) \rho_0 = \rho_0$ for $\rho_0$ in equilibrium. 
Very recently, expressions for higher order moments were also found for applications in qubit measurements and the parameter estimation of quantum system while no expressions for cumulants or
higher order spectra were derived  \cite{atalayaPRA2018,tilloyPRA2018} .

\begin{widetext}
\section{Multi-time cumulants $C_3$ and $C_4$}
\label{sec:multicum}
Here we derive expressions for the multi-time cumulants of $z(t)$ that are of great importance for describing actual measurements. 
Unlike moments, cumulants allow for a straightforward subtraction of background noise  that is statically independent from the actual quantum noise (like electronic noise in the measurement device, see Section \ref{sec:cumulants}).   
Moreover, a cumulant like $C_3(z(t_1),z(t_2),z(t_3))$ is strictly zero if any two of the arguments are uncorrelated. 
In contrast, the moment $\langle z(t_1)z(t_2)z(t_3) \rangle$ can still be non-zero although such a quantity is sometimes referred to as a "correlator" \cite{tilloyPRA2018}.
 
After some algebra we find a surprisingly simple way of expressing the cumulant that very much resembles 
the expressions for the multi-time moments that were given above. 
We first express $C_3(z(t_1),z(t_2),z(t_3))$ in terms of the multi-time moments, Eq.~(\ref{eq:MomentsK}), 
for time order $t_3 > t_2 > t_1$
\begin{eqnarray}
 \langle  z(t_1)\rangle &  & = \beta^2 {\rm Tr}[A \rho_0]   \nonumber \\ 
  \langle  z(t_2) z(t_1)\rangle & = &\beta^4 {\rm Tr}[A {\cal G}(t_2 - t_1){\cal A}\rho_0]  \nonumber \\
  \langle z(t_3) z(t_2) z(t_1)\rangle 
 & = & \beta^6 {\rm Tr}[A {\cal G}(t_3 - t_2){\cal A}{\cal G}(t_2 - t_1){\cal A}\rho_0].  \label{zTheoryMomentsC3}
\end{eqnarray}
Considering the third order cumulant
\begin{eqnarray}
 C_3(x_1,x_2,x_3) & = & \langle x_3 x_2 x_1\rangle \nonumber \\
  & & - \langle x_3\rangle \langle x_2 x_1\rangle - \langle x_2\rangle \langle x_3 x_1\rangle -
  \langle x_1\rangle \langle  x_2 x_1\rangle  \nonumber \\
& &  + 2 \langle x_3\rangle \langle x_2 \rangle \langle x_1\rangle \label{eq:thirdOrderCumulant}
\end{eqnarray}   
we find an expression for $C_3(z(t_1),z(t_2),z(t_3))$ in terms of the multi-time moments
\begin{eqnarray}
 \beta^{-6}C_3(z(t_1),z(t_2),z(t_3)) & = & \sum_{\rm prm. t_j} \left[  {\rm Tr}[A {\cal G}(t_3 - t_2){\cal A}{\cal G}(t_2 - t_1){\cal A}\rho_0]  
    - {\rm Tr}[A \rho_0] {\rm Tr}[A {\cal G}(t_2 - t_1){\cal A}\rho_0] + \frac{1}{3} {\rm Tr}[A \rho_0]^3 \right].
\end{eqnarray}    
The sum over all permutations and the property ${\cal G}(t) = 0$ for $t\le 0$ guarantees that all terms in Eq.~(\ref{eq:thirdOrderCumulant}) are correctly represented.
Moreover, the permutations also guarantee that the RHS of the above equation is not sensitive to the time order of $t_1$, $t_2$, and $t_3$. I.e. the equation for $C_3$ is correct for all time orders as long as $t_j \neq t_k$ holds for any pair $t_j$ and $t_k$ with different indices. The last term 
$\sum \frac{1}{3} {\rm Tr}[A \rho_0]^3$ can be found to correspond to $2 \langle x_3\rangle \langle x_2 \rangle \langle x_1\rangle$ if
all six permutations are taken into account. 

Next, we show that the expression for the cumulant $C_3$ dramatically simplifies to a single term under the sum after introducing
a new modified propagator
\begin{equation}
 {\cal G}'(t) = {\cal G}(t) - {\cal G}_0  \Theta(t) \label{eq:G'}
\end{equation}
where
\begin{equation}
{\cal G}_0 := \lim_{t' \rightarrow \infty}{\cal G}(t'). 
\end{equation}
The step function $\Theta(t) = 1$ for $t>0$ and zero otherwise guarantees that ${\cal G}'(t) = 0$ for $t\le0$. We find the important property
\begin{eqnarray}
  {\cal G}_0 X & = & {\rm lim}_{t \rightarrow \infty}{\cal G}(t) X \nonumber \\
  & = & \rho_0 {\rm Tr}(X) \label{eq:traceproperty}
\end{eqnarray}
since (i) ${\cal G}(t)$ for $t> 0$ always conserves the trace, (ii) the left hand side of the equation is always proportional to $\rho_0$, and (iii) ${\rm Tr}\rho_0 = 1$.
It follows that the superoperator ${\cal G}'(t)$ annihilates the steady state contribution $\rho_0$ of any operator to its right. 
The cumulant $C_3$ becomes after replacing ${\cal G}(\tau)$ by ${\cal G}'(\tau)+{\cal G}_0(\tau)  \Theta(\tau)$ and expansion
\begin{eqnarray}
 \beta^{-6} C_3 & = & \sum_{\text{prm. $t_j$}} \left[ {\rm Tr}[A {\cal G}'(t_3 - t_2){\cal A}{\cal G}'(t_2 - t_1){\cal A}\rho_0] \right. \nonumber \\
  & & + {\rm Tr}[A \rho_0] {\rm Tr}[A {\cal G}'(t_3 - t_2){\cal A}\rho_0]\Theta(t_2 - t_1) \nonumber  \\
   & & + {\rm Tr}[A \rho_0] {\rm Tr}[A {\cal G}'(t_2 - t_1){\cal A}\rho_0]\Theta(t_3 - t_2) \nonumber  \\ 
   & & +  {\rm Tr}[A \rho_0]^3 \Theta(t_2 - t_1)\Theta(t_3 - t_2) \nonumber \\
   & & \left. - {\rm Tr}[A \rho_0] {\rm Tr}[A {\cal G}' {\cal A}(t_2 - t_1)\rho_0]  - {\rm Tr}[A \rho_0]^3 \Theta(t_2-t_1) + \frac{1}{3} {\rm Tr}[A \rho_0]^3
   \right].
 \end{eqnarray}
The terms that are third order in $\rho_0$ cancel. All terms with the factor   ${\rm Tr}[A {\cal G}'(t_j - t_k){\cal A}\rho_0]$ can be written in terms of ${\cal G}'(t_2 - t_1)$ regarding that they all appear in a common sum over all permutations. 
We find
\begin{eqnarray}
 \beta^{-6}C_3 & = & \sum_{\text{perm. $t_j$}} \left[ {\rm Tr}[A {\cal G}'(t_3 - t_2){\cal A}{\cal G}'(t_2 - t_1){\cal A}\rho_0] \right. \nonumber \\
 & & + {\rm Tr}[A \rho_0] {\rm Tr}[A {\cal G}'(t_2 - t_1){\cal A}\rho_0](\Theta(t_1 - t_3) + \Theta(t_3-t_2) -1 ). 
\end{eqnarray}
The second term is obviously zero for $t_2 < t_1$ because of the factor $ {\cal G}'(t_2 - t_1)$. Moreover, the term in the braces makes the term zero except for $t_2>t_3>t_1$. Since $ {\cal G}'(t_2 - t_1) = {\cal G}'(t_2 - t_3){\cal G}'(t_3 - t_1)$ for $t_2>t_3>t_1$ and zero otherwise, we can write 
 \begin{eqnarray}
 \beta^{-6}C_3 & = & \sum_{\text{prm. $t_j$}}  \left[ {\rm Tr}[A {\cal G}'(t_3 - t_2){\cal A}{\cal G}'(t_2 - t_1){\cal A}\rho_0] \right. \nonumber \\
 & & \left. - {\rm Tr}[A \rho_0] {\rm Tr}[A {\cal G}'(t_2 - t_3){\cal G}'(t_3 - t_1){\cal A}\rho_0] \right] \\
 & = &  \sum_{\text{prm. $t_j$}}  {\rm Tr}[{\cal A} {\cal G}'(t_3 - t_2)({\cal A}- {\rm Tr}[A\rho_0]){\cal G}'(t_2 - t_1){\cal A}\rho_0] 
\end{eqnarray}
where we permuted the indices in the second term to obtain the same structure as in the first term. 
The first factor $A$ in the last line was replaced by ${\cal A}$ due to the cyclicity of the trace. The term $({\cal A}- {\rm Tr}[A\rho_0])$
 seems somewhat to disturb the symmetry in the last line. The symmetry is made perfect by noting that 
the last factor ${\cal A}$ can be replaced by ${\cal A} - {\rm Tr}[A \rho_0]$ since ${\cal G'}(t) \rho_0 = {\cal G}(t) \rho_0 - {\cal G}_0 \rho_0
= \rho_0 - \rho_0$ is always zero [see Eq.~(\ref{eq:G'})].
The first factor  ${\cal A}$ can also be replaced by ${\cal A} - {\rm Tr}[A \rho_0]$ since ${\rm Tr}{\cal G}'(t) X $ is always zero as  
follows from the trace property above [Eq.~(\ref{eq:traceproperty})].
After introducing ${\cal A}' = {\cal A} - {\rm Tr}[A \rho_0]$ we eventually find 
 \begin{eqnarray}
 C_3(z(t_1),z(t_2),z(t_3)) & = & \beta^6 \sum_{\text{prm. $t_j$}}  {\rm Tr}[{\cal A}' {\cal G}'(t_3 - t_2){\cal A}'{\cal G}'(t_2 - t_1){\cal A}'\rho_0]. 
 \label{eq:C3}
\end{eqnarray}
The structure of the cumulant is as simple as that of the corresponding moment $M_3$
where the operator ${\cal G}$ in Eq.~(\ref{eq:MomentsKperm}) is replaced by ${\cal G'}$ and ${\cal A}$ by ${\cal A}'$ to yield Eq.~(\ref{eq:C3}). This is very remarkable considering that $C_3$ represents a cumulant whose initial structure, Eq.~(\ref{eq:thirdOrderCumulant}), contains no less than 5 terms. Since ${\cal G}'(t)$ always decays to zero for large $t$, we immediately find that also $C_3$ decays to zero for large time differences. Such a behavior was anticipated for cumulants of $z(t)$ in Section \ref{sec:cumulants}. 
The very simple structure of the $C_3$ result suggest that a similar structure also holds for the fourth order case. Starting from a sum of initially 15 terms we indeed were able to derive the following simple representation of the cumulant (see Appendix \ref{AppC4})
 \begin{eqnarray}
 C_4(z(t_1),z(t_2),z(t_3),z(t_4)) & = & \beta^8 \sum_{\text{prm. $t_j$}} {\rm Tr}[{\cal A}' {\cal G}'(t_4 - t_3){\cal A}'{\cal G}'(t_3 - t_2){\cal A}'{\cal G}'(t_2 - t_1){\cal A}'\rho_0]. \label{eq:C4}
\end{eqnarray}
 We consider the cumulant expressions Eqs.~(\ref{eq:C3}) and (\ref{eq:C4}) a major simplification in the field of higher order quantum noise. Previously, expressions for $C_3$ kept the
  unwieldy structure of Eq.~(\ref{eq:thirdOrderCumulant}), see e.g. Eq.~(30) in Ref. \onlinecite{marcosNJP2010}. So far we were, however, not able to find a proof for a general cumulant formula beyond $C_4$.
\section{The trispectrum and bispectrum}
\label{sec:Trispectrum}
The multi-time cumulant $C_4$ of $z(t)$ found above can in principle be calculated numerically. A subsequent numerical three-dimensional Fourier transform would then result in the trispectrum $S_z^{(4)}(\omega_1,\omega_2,\omega_3)$
 [see Eq.~(\ref{DefPolyspectra})]. This procedure has the disadvantage that a three dimensional space of data points has to be evaluated before even a single point of the spectrum $S_z^{(4)}$ can be calculated. 
We therefore derive in the following an explicit quantum mechanical expression in the frequency domain that allows for a point-wise evaluation of $S_z^{(4)}$ at a given position $\omega_1,\omega_2,\omega_3$. This tremendously saves computing time in cases where e.g. only a two-dimensional cut through $S_z^{(4)}$ is of interest or when high frequencies would require a very fine sampling of $C_4$ in the time domain. 

The Fourier transformation of $C_4(z(t_1),z(t_2),z(t_3),z(t_4))$ faces the problem that we have no expression for $C_4$ for the cases of equal times (e.g. $t_2 = t_3$). We will show in the next paragraph that we do not require that knowledge for getting a correct result for
  $C_4(z(\omega_1),z(\omega_2),z(\omega_3),z(\omega_4))$. 
First we note that
$C_n(z(\omega_n), z(\omega_{n-1}),\cdots,z(\omega_1))$ decays for $n \ge 3$ to zero for any $\omega_j \rightarrow \pm \infty$.
This property can be shown by considering the detector output Eq.~(\ref{SME_detector}) 
in the Fourier domain
\begin{equation}
 z(\omega) = \beta^2 {\rm Tr}(\rho(\omega) A) +\beta \frac{1}{2} \Gamma(\omega).
\label{detectorOmega}
  \end{equation}
The first term is related to system dynamics $\rho(t)$ and decays quickly to zero for frequencies outside the interval of frequencies of system resonances (compare Fig.~\ref{spinNoiseSpectrum}).
After using Eq.~(\ref{detectorOmega}) and the multi-linearity of cumulants to decompose $C_n$ into a sum  \cite{nikiasIEEE1993}, all terms [like e.g. $C_n(\beta\Gamma(\omega_n)/2, \beta^2 {\rm Tr}(\rho(\omega_{n-1}) A), \cdots)$] except 
$C_{\rm remainder} = C_n(\beta\Gamma(\omega_n)/2,\cdots, \beta\Gamma(\omega_1)/2)$ will decay to zero for any $\omega_j \rightarrow \pm \infty$. Also, $C_{\rm remainder} = 0$ for $n \ge 3$ since $\Gamma(\omega_j)$ are purely Gaussian. Only for $n=2$
we find $C_2(z(\omega_2),z(\omega_1)) \rightarrow 2\pi \delta(\omega_1 + \omega_2)\beta^2/4 $ for any $\omega_j \rightarrow \pm \infty$ 
which easily follows from Eq.~(\ref{O6G}). 
The cumulant $C_4(z(t_4),z(t_3),z(t_2),z(t_1))$ was obtained above 
for the time order $t_4>t_3>t_2>t_1$. For other time orders like $t_4>t_2>t_3>t_1$ the cumulant is obtained 
from the same expressions after exchanging the corresponding indices ($t_2$ and $t_3$ in our example) on 
the right hand side of Eq.~(\ref{eq:MomentsK}) which effectively restores the required time-order.
Merely values of $C_4(z(t_4),z(t_3),z(t_2),z(t_1))$,  where one or more pairs of times $t_j$ are 
equal, are not defined via Eq.~(\ref{eq:MomentsK}) since a strict time order  $t_4>t_2>t_3>t_1$ {\it without} equal times was required for its derivation. We will abbreviate the time quadruple with the vector symbol $\vec{t}$ and relate to the set of vectors $\vec{t}$ with one or more pairs of equal time as $R_{\rm eq}$ and the set of all vectors as $R$.
In the following we will show that the calculation of $C_4(z(\omega_4), z(\omega_3),z(\omega_2),z(\omega_1))$ does only require the Fourier integral about $R \setminus R_{\rm eq}$.  The  Fourier integral about $R$ is four dimensional while the integral about $R_{\rm eq}$ is at most three dimensional. Consequently the $R_{\rm eq}$ contribution to the integral about $R$ is of order ${\rm d}t$ and may be fully neglected as long as $C_4(z(t_4),z(t_3),z(t_2),z(t_1))$ is finite on $R_{\rm eq}$. Suppose $C_4(z(t_4),z(t_3),z(t_2),z(t_1))$ had a $\delta$ singularity on $R_{\rm eq}$. Such a singularity would result in a non-vanishing contribution to $C_4(z(\omega_4), z(\omega_3),z(\omega_2),z(\omega_1))$ that would extend to infinite values of at least one $\omega_j$. Above we however showed that $C_4(z(\omega_4), z(\omega_3),z(\omega_2),z(\omega_1))$ tends to zero for any $\omega_j  \rightarrow \pm \infty$. Consequently, $\delta$-singularities are absent in $C_4(z(t_4),z(t_3),z(t_2),z(t_1))$ and  $C_4(z(\omega_4), z(\omega_3),z(\omega_2),z(\omega_1))$ can without any restrictions be calculated from the moments derived above without the need for deriving expression for $C_4(z(t_4),z(t_3),z(t_2),z(t_1))$ on $R_{\rm eq}$.

The following derivation of $S_z^{\rm (4)}(\omega_1,\omega_2,\omega_3)$ from $C_4(z(t_4),z(t_3),z(t_2),z(t_1))$ is based on that insight. 
The cumulant $C_4(z(\omega_1),z(\omega_2),z(\omega_3),z(\omega_4))$ can be expressed as a four dimensional Fourier transformation
\begin{equation}
  C_4(z(\omega_1),z(\omega_2),z(\omega_3),z(\omega_4))  
= \int \left(\sum_{\text{prm. $t_j$}} f(t_1,t_2,t_3,t_4) \right)e^{i \vec{\omega}^\intercal \vec{t}} {\rm d}^4\vec{t}
\end{equation}
where we abbreviated the term in the sum of Eq.~(\ref{eq:C4}) by $f$.
Instead of permuting the variables $t_j$, the variables $\omega_j$ can be permuted yielding
\begin{equation}
  C_4(z(\omega_1),z(\omega_2),z(\omega_3),z(\omega_4))  
= \sum_{\text{prm. $\omega_j$}} \int  f(t_1,t_2,t_3,t_4) e^{i \vec{\omega}^\intercal \vec{t}} {\rm d}^4\vec{t}.
\end{equation}
Considering that $f$ depends only on time differences, we can define a function $g$ with 
\begin{equation}
 g(t_2 - t_1, t_3 - t_2, t_4 - t_3) =  f(t_1,t_2,t_3,t_4) 
\end{equation} 
that is only three dimensional.
After introducing a corresponding transformation of the time variables
\begin{equation}
\left( \begin{array}{c} \tau_0 \\ \tau_1 \\ \tau_2 \\ \tau_3 \end{array} \right) = 
 \underbrace{\left( \begin{array}{cccc}  1 & 0 & 0 & 0 \\-1 & 1 & 0 & 0 \\ 0 & -1 & 1 & 0 \\ 0 & 0 & -1 & 1  \end{array} \right) }_B
 \left( \begin{array}{c} t_1 \\ t_2 \\ t_3 \\ t_4 \end{array} \right) 
\end{equation}
where ${\rm det} B = 1$ and
\begin{equation}
 B^{-1} = \left( \begin{array}{cccc} 1 & 0 & 0 & 0 \\ 1 & 1 & 0 & 0 \\ 1 & 1 & 1 & 0 \\ 1 & 1 & 1 & 1 \end{array} \right),
\end{equation}
we find
\begin{equation}
   C_4(z(\omega_1),z(\omega_2),z(\omega_3),z(\omega_4)) = \sum_{\text{prm. $\omega_j$}} \int g(\tau_1,\tau_2,\tau_3)  \exp (i \vec{\omega}^\intercal B^{-1} \vec{\tau}  ) {\rm d}^4\vec{\tau}.
\end{equation}
Introducing $\tilde{g}(\nu_1,\nu_2,\nu_3)$ as the three-dimensional Fourier transformation of $g(\tau_1,\tau_2,\tau_3)$
we find with $\vec{\nu} = (B^{-1})^\intercal \vec{\omega}$
\begin{eqnarray}
 C_4(z(\omega_1),z(\omega_2),z(\omega_3),z(\omega_4))& = &\sum_{\text{prm. $\omega_j$}} \tilde{g}(\omega_2 + \omega_3 + \omega_4, 
 \omega_3 + \omega_4,  \omega_4)  2\pi \delta(\omega_1+\omega_2+\omega_3+\omega_4).
\end{eqnarray}
This leads to a compact expression for the trispectrum 
\begin{eqnarray}
  S_z^{\rm (4)}(\omega_1,\omega_2,\omega_3,\omega_4 = -\omega_1-\omega_2-\omega_3) & = &  \sum_{\text{prm. $\omega_1,\omega_2,\omega_3,\omega_4$}}
 \hspace{-8mm}\beta^8{\rm Tr}[{\cal A}'{\cal G}'(\omega_4){\cal A}' {\cal G}'(\omega_3 + \omega_4){\cal A}'{\cal G}'(\omega_2 + \omega_3 + \omega_4){\cal A}'\rho_0].
 \label{eq:S4}
\end{eqnarray}
The bispectrum $S_z^{\rm (3)}$ is obtained from $C_3$ by a corresponding calculation as
\begin{eqnarray}
  S_z^{\rm (3)}(\omega_1,\omega_2,\omega_3 = -\omega_1-\omega_2) & = & 
   \sum_{\text{prm. $\omega_1,\omega_2,\omega_3$}}
  \beta^6{\rm Tr}[{\cal A}'{\cal G}'(\omega_3){\cal A}'{\cal G}'(\omega_3 + \omega_2){\cal A}'\rho_0]. \label{eq:S3}
\end{eqnarray}
The usual power spectrum $S_z^{\rm (2)}(\omega)$ is given by
\begin{equation}
  S_z^{(2)}(\omega) = \beta^4 ( {\rm Tr}[{\cal A}'{\cal G}'(\omega){\cal A}'\rho_0] +  {\rm Tr}[{\cal A}'{\cal G}'(-\omega){\cal A}'\rho_0] 
  ) + \beta^2/4. \label{eq:S2}
\end{equation}
where the additional last term follows from the calculation of $M_2 = \langle z(t_2) z(t_1) \rangle$ if the case $t_2 = t_1$ is included.
We note that the fourth order in $\beta$ contribution to $S_z^{(2)}(\omega)$ 
can be transformed into $\beta^4 S_{\rm q}(\omega)$
[see Eq. (\ref{spectrumQM})].
We emphasize that the expressions for the spectra $S_z^{(2)}$,  $S_z^{(3)}$, and  $S_z^{(4)}$ are free from $\delta$-function contributions since 
${\cal G}'(\tau)$ does unlike  ${\cal G}(\tau)$ no longer contain the constant contribution ${\cal G}_0$. The numerical treatment of the spectra via ${\cal G}'(\omega)$ - see Eq.~(\ref{eq:NumericsG'}) -  faces no trouble with $\delta$-functions.  
\end{widetext}

\section{Higher order spectra of a coupled spin pair}    
\label{sec:S4spinpair}
\begin{figure*}
   \centering
 \includegraphics[width=16cm]{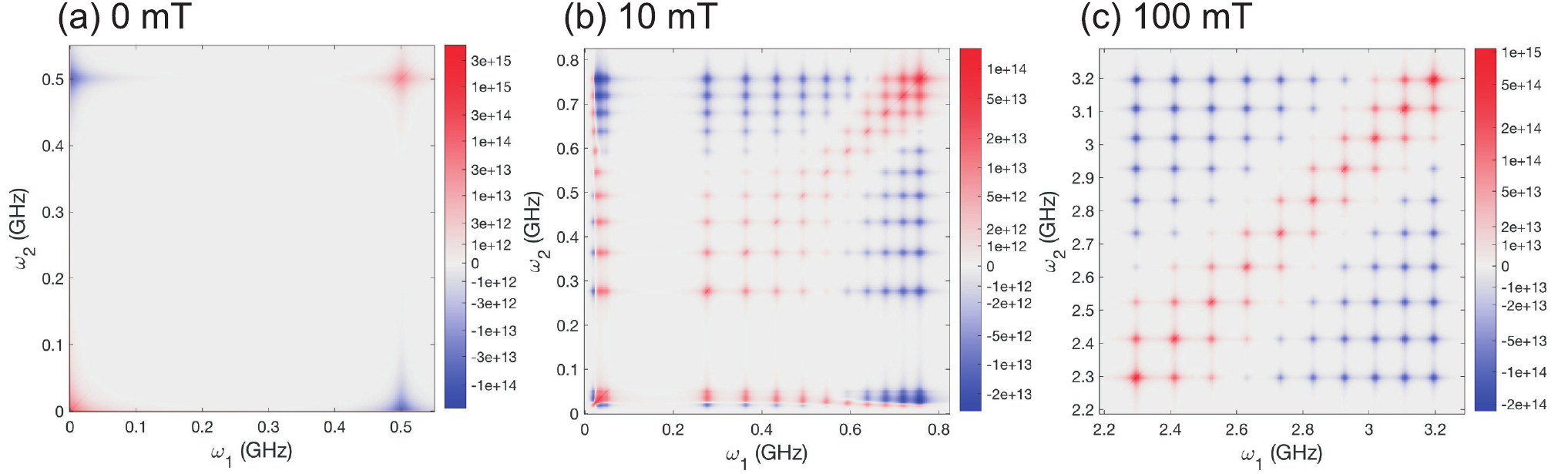} 
   \caption{ Fourth order   correlation spectra $S_z^{(4)}(\omega_1,-\omega_1,\omega_2,-\omega_2)$ of the spin-spin system in ZnO:In at $B_x = 0$, $10$, and $100$~mT. Negative correlations appear in blue. The function asinh is used for color-scaling.}
\label{CorrelationSpectrum}
\end{figure*}
\begin{figure}
   \centering
 \includegraphics[width=9cm]{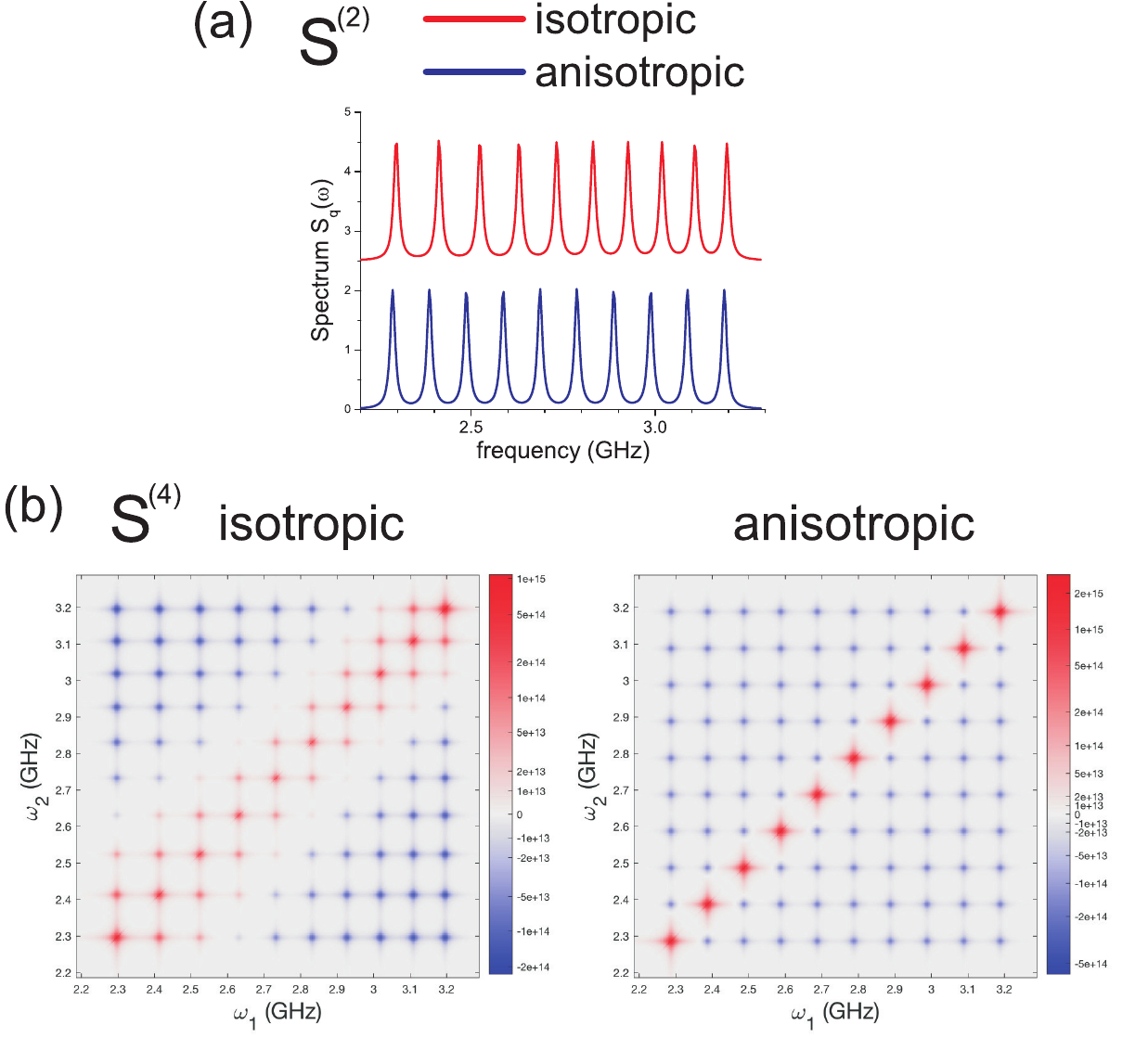} 
   \caption{Comparison of spin noise spectra of the coupled spin system in ZnO:In at $B_x = 100$~mT for isotropic and anisotropic hyperfine-interaction along the $x$-axes. (a) Second order spectra $S_{\rm q}(\omega)$. (b) Fourth order correlation spectra $S_z^{(4)}(\omega_1,-\omega_1,\omega_2,-\omega_2)$. }
    \label{CorrelationSpectrumAniso}
    \end{figure}
        \begin{figure*}
   \centering
 \includegraphics[width=18cm]{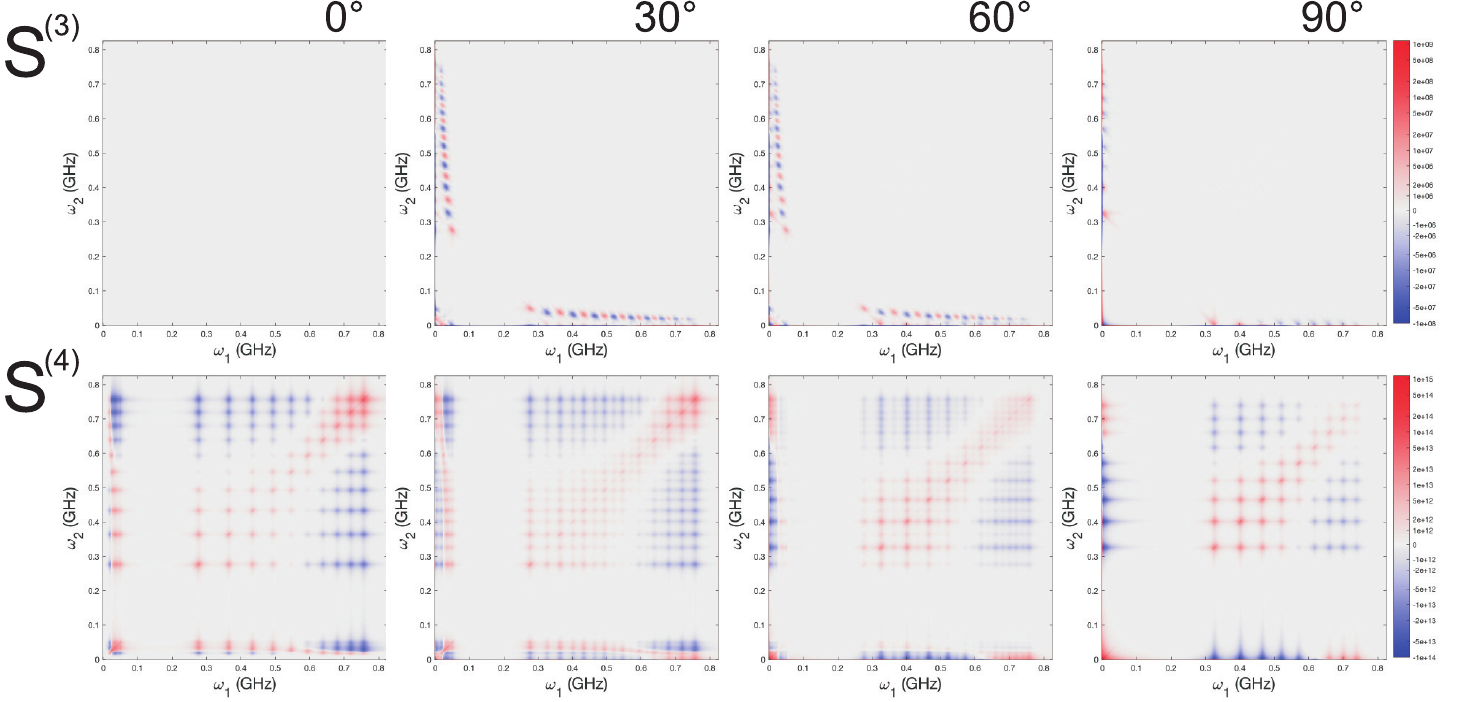} 
   \caption{ Bispectrum $S_z^{(3)}(\omega_1,\omega_2,-\omega_1 -\omega_2)$ (upper row) and correlation spectrum $S_z^{(4)}(\omega_1,-\omega_1,\omega_2, -\omega_2)$ (lower row) for a 10~mT magnetic field 
   in in-plane ($0^\circ$) and out-off-plane directions ($30^\circ,\,60^\circ,\,90^\circ$).}
\label{figureS3S4}
\end{figure*}
Figure \ref{CorrelationSpectrum} 
shows cuts $S_z^{(4)}(\omega_1,-\omega_1,\omega_2,-\omega_2)$ through the trispectrum for the ZnO spin-spin system at $T = 10$~K discussed in Section \ref{sec:Examples} for different magnetic fields in $x$-direction. The two-dimensional cuts are identical with the definition of the correlation spectrum Eq.~(\ref{eq:fourthOrderCorrelation}) provided $\omega_1 \neq \omega_2$ and $\omega_1,\omega_2 \neq 0$. 
The cuts have been computed numerically via MATLAB using Eq.~(\ref{eq:S4}) where the superoperator ${\cal L}$ was represented as a matrix $L$.
After numerically computing the matrix of eigenvectors $\Lambda$ of $L$ we can write 
\begin{equation}
 L = \Lambda D_L \Lambda^{-1}
 \end{equation}
where $D$ is a diagonal matrix of the eigenvalues $\lambda_j$.  The superoperator ${\cal G}(t)$ ($t > 0$) can thus be 
represented by
\begin{equation}
 G(t) = \Lambda D_G \Lambda^{-1}
\end{equation}
where $D_G$ contains the diagonal elements $\exp (\lambda_j t)$.
For a damped quantum system we find that one eigenvalue $\lambda_m$ is zero with corresponding eigenvector $r_0$  that represents the steady state $\rho_0$. All other eigenvalues $\lambda_j$ ($j \neq m$) have a negative real part as ${\cal G}(t)$ always forces the system into the steady state. The diagonal entry $\exp (\lambda_m t) = 1$ is related to ${\cal G}_0 = {\cal G}(\infty)$. The superoperator ${\cal G}'(t)$ [Eq.~(\ref{eq:G'})] that we introduced for the calculation of cumulants  is consequently given very simply as
\begin{equation}
 G'(t) = \Lambda D_{G'} \Lambda^{-1}
\end{equation}
where $D_{G'}$ contains the diagonal elements $\exp (\lambda_j t)$ for $j \neq m$ and is zero for $j = m$.
The Fourier transform ${\cal G}'(\nu)$ is found to be
\begin{equation}
 G'(\nu) = \Lambda D_{\tilde{G}'} \Lambda^{-1} \label{eq:NumericsG'}
\end{equation}
where $D_{\tilde{G}'}$ contains the diagonal elements $1/(-\lambda_j  - i\nu)$ for $j \neq m$ and  zero for $j = m$.
The bispectrum und trispectrum are then easily computed directly from Eqs.~(\ref{eq:S3}) and  (\ref{eq:S4}).
The correlation spectrum for 100~mT [Figure \ref{CorrelationSpectrum}(c)] exhibits both positive and negative correlations. Positive correlation appear among spectral lines that are close in frequency while negative correlations appear for lines that are distant in frequency. 
The spectrum can be interpreted in the following way after noting that (i) the total angular moment $I_x + s_x$ in B-field direction is conserved by the Hamiltonian apart from the small contribution proportional to $P_{\parallel}$; (ii) the spectral peaks are due to the electron spin precessing in an effective field which is a sum of the hyperfine field and the external magnetic field. System states with large $I_x+s_x$ values correspond to low frequency peaks while nuclear states with negative $I_x+s_x$ values result in the high frequency peaks. 
If now $z(t)$ reveals some spectral weight at one of the peak frequencies, the system reveals also some information on its approximate total angular momentum (the angular momentum is not exactly revealed because the signal is due to the beating of at least two eigenstates that not necessarily have the same total angular momentum). Consequently, $z(t)$ will with an increased probability contain spectral weight at frequencies that are compatible with that total angular momentum. Since $H$ conserves the total angular momentum in $x$-direction, only damping will force the system with time into states that belong to different total angular momenta. Consequently, the simultaneous appearance of spectral weight at distant frequencies is suppressed leading to negative values in the correlation spectrum.
For absent field [$B = 0$, Figure \ref{CorrelationSpectrum}(a)] we find spectral weight at zero which we interpret as an electron and a nucleus that are aligned parallel or anti-parallel to the direction of observation ($z$) giving rise to a constant offset signal. The offset disappears when the electron is precessing around an axes in the $xy$-plane (given by the nuclear spin) resulting in the frequency peak around 0.5~GHz. The two situations exclude each other which gives rise to the negative off-diagonal peak. Towards $B = 10$~mT we find a transition to a strongly structured spectrum that we currently are not able to interpret.

    \begin{figure}
   \centering
 \includegraphics[width=6cm]{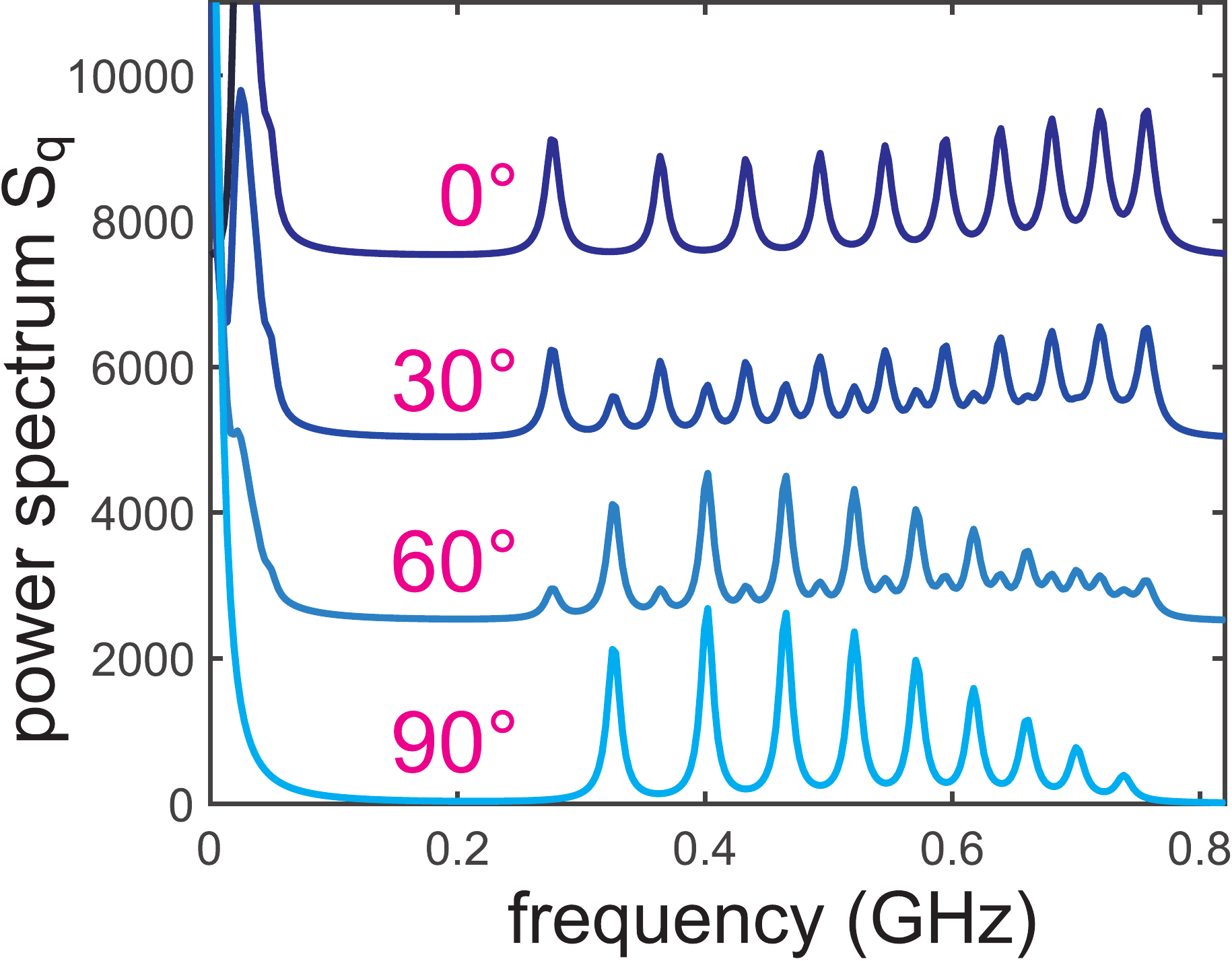} 
   \caption{Power spectra $S_{\rm q}(\omega)$ for a 10~mT magnetic field in in-plane ($0^\circ$) and out-off-plane directions ($30^\circ,\,60^\circ,\,90^\circ$).}
\label{figureS2winkel}
\end{figure}
The correlation spectrum is found to sensitively depend on the tensor of the hyperfine interaction. Figure \ref{CorrelationSpectrumAniso}(b) compares the correlation spectra for the system at 100~mT for the isotropic hyperfine interaction $H_{\rm hyp} = A\vec{I} \cdot \vec{s}$ with anisotropic hyperfine interaction in $x$-direction only, i.e. $H_{\rm hyp} =  A I_x s_x$. All frequency peaks become anti-correlated
in the case of anisotropic interaction. The Hamiltonian conserves now $I_x+s_x$, $I_x$, and $s_x$ at the same time. A certain electron precession frequency belongs therefore to a well defined value of $I_x$. Since the appearance of one frequency reveals the system to be in a certain $I_x$-nuclear state it can not at the same time exhibit another frequency that would belong to another $I_x$-nuclear state. Consequently all frequencies must be anticorrelated as revealed by the correlation spectrum. A comparison between the usual power spectrum [Figure  \ref{CorrelationSpectrumAniso}(a)] exhibits only slight changes in the absolute peak positions but no overall change of the structure. 

Correlation spectra clearly reveal additional information compared to $S_z^{(2)}$ [Figure \ref{spinNoiseSpectrum}]. 
For large magnetic fields the $S_z^{(2)}$ spectrum exhibits ten separate peaks. A quantum system consisting of
 ten independent electron spins precessing at ten different frequencies (due to e.g.  material dependent g-factors) 
 could display exactly the same structure of ten peaks in $S_z^{(2)}$.  The higher order spectrum $S_z^{(4)}$ would 
 however exhibit no cross-correlations contributions while the higher order spectrum of the coupled spin-spin
  system exhibits strong positive and negative correlations.

Last, we directly compare bispectra $S_z^{(3)}(\omega_1,\omega_2,-\omega_1 -\omega_2)$ of the ZnO:In system with correlation spectra $S_z^{(4)}(\omega_1,-\omega_1,\omega_2, -\omega_2)$ for a magnetic field of $|\vec{B}| = 10$~mT pointing at different angles $\varphi$ out of the $xy$-plane (Figure \ref{figureS3S4}). Figure
 \ref{figureS2winkel} shows for comparison the corresponding power spectra $S_{\rm q}(\omega)$. 
The imaginary part of $S_z^{(3)}$ is zero for all angles which is consistent with time-reversal symmetry of $z(t)$ 
for a system in thermal equilibrium \cite{balkPRX2018}. Interestingly, also the real part of  $S_z^{(3)}$ is zero for $\varphi = 0$ 
while $S_z^{(4)}$ exhibits strong correlations for all angles. The bispectrum $S_z^{(3)}$ displays positive and 
negative values for angles of 30, 60, and 90$^\circ$. Their appearance is, however, limited to regions 
close to the $\omega_1$- and $\omega_2$-axes. We found that $S_z^{(3)}$ disappears completely for higher magnetic fields: 
Since $S_z^{(3)}$ arises from $C_3(z(\omega_1), z(\omega_2), z(-\omega_1-\omega_2))$ a non-zero cumulant $C_3$ 
requires at least simultaneously spectral weight of $z(\omega)$ at {\it three} frequencies $\omega_1$, $\omega_2$, and $-\omega_1-\omega_2$. The spectral weight of the system dynamics is for large magnetic fields given by the ten peaks centered 
around $\omega = \pm \beta g^{(e)} B /\hbar$. For $\omega_1 \approx \omega$ and $\omega_2 \approx \omega$ we
 find $-\omega_1 - \omega_2 \approx -2 \omega$. However, the spin-spin system exhibits no spectral weight 
 at $\pm 2\omega$ which implies that $C_3$ and the bispectrum $S_z^{(3)}(\omega_1,\omega_2)$  disappear. 
 Similarly, $\omega_1 \approx \omega$ and $\omega_2 \approx -\omega$ also yield no contribution to $S_z^{(3)}$.
Non-zero contributions to $S^{(4)}$ correlation spectra require only spectral weight at {\it two} frequencies 
 $\omega_1$ and $\omega_2$  (implying also weight at $-\omega_1$ and $-\omega_2$) which explains their much richer structure compared to $S^{(3)}$. The example above clearly shows that a bispectrum $S^{(3)}$ may sometimes be blind to correlations in the system dynamics whose presence, however, is revealed in the correlation spectrum $S^{(4)}$.  
 
The question arises of what information about system parameters can be obtained from spectra
  $S^{(2)}$, $S^{(3)}$, $S^{(4)}$ or higher. Without going into details it can be said that in the absence 
  of damping the energy levels $E_j$ of the quantum system, the transition matrix elements $\langle j| A | k \rangle$, 
  and the temperature $T$ (which gives $\rho_0$) are sufficient to calculate all higher order spectra $S^{(n)}$. 
  Quantum beats between states $j$ and $k$ with respect to a transition induced by $A$ can give rise to the appearance 
  of spectral peaks at frequencies $\omega = (E_j- E_k)/\hbar$. A number of $m$ peaks in a spectrum $S^{(2)}$ could
   be explained by the beats between $m$ pairs of otherwise uncorrelated quantum states. Taking into account also
    the spectrum $S^{(4)}$, a positive correlation of frequencies at $S^{(4)}(\omega_1,\omega_2)$ would indicate 
    that two of the pairs share a common quantum states. 
We expect that the new availability of a quantum expression for higher order spectra will trigger
 further research on the reconstruction of system parameters. The theory of continuous matrix 
 product states (cMPS) was recently identified as a possible
 very general basis for going into that direction \cite{tilloyPRA2018, hubenerPRL2013}.

\section{conclusion}
In conclusion, we presented quantum mechanical expressions for the calculation of higher order moments, cumulants, 
and spectra of the detector output $z(t)$ of a continuously measured quantum systems. All expressions are given in
 terms of the system propagator ${\cal G}(\tau)$, the measurement operator ${\cal A}$, and the steady state density 
 matrix $\rho_0$. The simple structure of the expressions for the moments, Eq.~(\ref{eq:MomentsK}), 
 could (for $n=3$ and $n=4$) surprisingly be shown to reappear also for the cumulants, Eq.~(\ref{eq:C4}), after introduction of modified 
superoperators ${\cal G}'$ and ${\cal A}'$. Eventually, we presented compact expressions for the
  power spectrum $S_z^{(2)}$, the bispectrum $S_z^{(3)}$, and the tripsectrum $S_z^{(4)}$, Eqs.~(\ref{eq:S2}), (\ref{eq:S3}), and (\ref{eq:S4}). While $S_z^{(2)}$ has appeared in many versions in the literature before, we are not aware of any previous
   general expressions for $S_z^{(3)}$ and $S_z^{(4)}$. The new expressions are valid for continuous measurements
  of arbitrary strength including the Zeno-limit and allow for a treatment of external damping in Markov-approximation. 
 The new expressions therefore cover applications in spin noise spectroscopy where weak measurements are realized 
 (with a large background of Gaussian noise) as well as transport measurements where usually the limit of strong
 measurements holds (quantum jumps and telegraph noise). We expect future application of our expressions in
 quantum optics, transport theory, quantum information science, and measurement theory in general. 
 Future extensions of our theory may include cumulant expressions for the simultaneous measurement of more
 than one observable or the description of noise of coherently driven systems \cite{jacobsCP2006,forguesSR2013}.
\begin{acknowledgments}
We thank R. Grauer, J. K\"onig, P. Stegmann, and F. Anders for stimulating discussions and acknowledge financial support of the DFG under Grant No. HA 3003/7-1.
\end{acknowledgments}
\appendix
\section{Conventions for Fourier transformations and Convolution Integrals}
\label{app:Conventions}
We distinguish a function $f(t)$ and its Fourier transform $f(\omega)$ only by its argument. They are related by
\begin{eqnarray}
 f(\omega) & = & \int_{-\infty}^{\infty} e^{i \omega t}f(t)\,{\rm d}t \\
  f(t) & = & \frac{1}{2\pi}\int_{-\infty}^{\infty} e^{-i \omega t}f(\omega)\,{\rm d}\omega.
\end{eqnarray} 
We define the convolution in time as
\begin{equation}
 f(t)\ast g(t) = \int_{-\infty}^{\infty} f(t-\tau)g(\tau)\,{\rm d}\tau
\end{equation}
and the convolution in frequency as
\begin{equation}
  f(\omega) \ast g(\omega) = \frac{1}{2\pi}\int_{-\infty}^{\infty} f(\omega-\nu)g(\nu)\,{\rm d}\nu
\end{equation}
with the additional prefactor $(2 \pi)^{-1}$.
This leads to the following relations for the Fourier transforms of convolutions and products of functions:
\begin{eqnarray}
 h_1(t)&= &f(t) \ast g(t) \\
 h_1(\omega) & = & f(\omega)g(\omega)
\end{eqnarray}
and
\begin{eqnarray}
 h_2(\omega) & = & f(\omega)*g(\omega) \\
 h_2(t) & = & f(t)g(t).
\end{eqnarray}
The above relations also hold if $f$ and $g$ are operators as long as the operator ordering is kept the same during the operations.
\section{Moments of white noise}
\label{app:MomentsAndCumulants}
Real valued white Gaussian noise $\Gamma(t)$ fulfils the relation $\langle\Gamma(t) \Gamma(t')\rangle = \delta(t-t')$. Higher order moments of $\Gamma(t_i)$ are given by Isserlis' theorem:
\begin{eqnarray}
 \langle \Gamma(t_1) \Gamma(t_2) \Gamma(t_3) \Gamma(t_4)\rangle \hspace{-3cm} & & \nonumber \\
  & = &  \delta(t_1 - t_2) \delta(t_3 - t_4) \nonumber \\
  & & + \delta(t_1 - t_3) \delta(t_2- t_4) \nonumber \\
  & &   + \delta(t_1 - t_4) \delta(t_2- t_3) .
\end{eqnarray}
The $n$th order of even $n$ can be recursively calculated via
\begin{eqnarray}
\langle \Gamma(t_1) \Gamma(t_2) ... \Gamma(t_n)\rangle \hspace{-4cm} & & \nonumber \\
  & = &  \delta(t_1 - t_2)  \langle \Gamma(t_3) \Gamma(t_4) ... \Gamma(t_n)\rangle \nonumber \\
  & & + \delta(t_1-t_3) \langle \Gamma(t_2) \Gamma(t_4) ... \Gamma(t_n)\rangle \nonumber  \\
    & & + \delta(t_1 - t_4) \langle \Gamma(t_2)\Gamma(t_3) \Gamma(t_5) ... \Gamma(t_n)\rangle  \nonumber \\
  & &   + \delta(t_1 -t_5) \langle \Gamma(t_2) .. \Gamma(t_4) \Gamma(t_6) ... \Gamma(t_n)\rangle \nonumber \\
  & & + ... \nonumber \\
  & & + \delta(t_1 - t_n) \langle \Gamma(t_2) ... \Gamma(t_{n-1})\rangle.
\end{eqnarray}
All odd-order moments are zero.
\begin{widetext}
\section{Equivalence of $G_{\rm q}(\tau)$ and $G^{\rm (L)}(\tau)$ for absent damping}
\label{app:Equivalence}
Here, we show that $G_{\rm q}(\tau)$ and $G^{\rm (L)}(\tau)$ coincide for absent damping and $\langle A(0)\rangle = 0$.
We use $A(\tau) = e^{i H \tau/\hbar} A(0) e^{-i H \tau/\hbar}$ and find
\begin{eqnarray}
 \langle A(-\tau) A(0) \rangle & = & {\rm Tr}( A(-\tau) A(0) \rho_0) \nonumber \\
  & = &  {\rm Tr}( A(-\tau) e^{-i H \tau/\hbar} e^{i H \tau/\hbar}A(0) e^{-i H \tau/\hbar} e^{i H \tau/\hbar}\rho_0 e^{-i H \tau/\hbar} e^{i H \tau/\hbar}) \nonumber \\
    & = &  {\rm Tr}(e^{i H \tau/\hbar} A(-\tau) e^{-i H \tau/\hbar} e^{i H \tau/\hbar}A(0) e^{-i H \tau/\hbar} e^{i H \tau/\hbar}\rho_0 e^{-i H \tau/\hbar} ) \nonumber \\
    & = & {\rm Tr}( A(0) A(\tau) \rho_0) \nonumber \\
     & = & \langle A(0) A(\tau) \rangle  \label{COMP1}
    \end{eqnarray}
where we used the fact  that the trace is conserved for cyclic permutations of the operators (line three) and that $\rho_0$ does not change with time when in thermal equilibrium (line four). 
\end{widetext}
The above equation also implies $\langle A(0) A(-\tau) \rangle  = \langle A(\tau) A(0) \rangle $.
Consequently we find for $\langle A(\tau)\rangle = \langle A(0)\rangle = 0$ that 
\begin{eqnarray}
 G^{\rm (L)}(\tau) & = & \frac{1}{2}\langle A(\tau) A(0) + A(0) A(\tau) \rangle \nonumber \\
  & = & \frac{1}{2}\langle A(-\tau) A(0) + A(0) A(-\tau) \rangle \nonumber \\
  & = & \frac{1}{2}\langle A(|\tau|) A(0) + A(0) A(|\tau|) \rangle \nonumber \\
   & = & G_{\rm q}(\tau).
\end{eqnarray} 
\section{Ito treatment of the SME and CQNF}
\label{sec:Ito}
Here, we employ Ito-calculus to solve Eq.~(\ref{SME_approx}). Consider the rewritten stochastic differential equation
\begin{equation}
  {\rm d}x(t) = {\cal L}x(t) \,{\rm d}t+  a(t)\,{\rm d}W(t)  \label{SMEiteration}
\end{equation}
with $x(t) = \rho_{n+1}$ and $a(t) = \lambda \beta B_n$ where $a(t)$ is a matrix-valued stochastic process
that is uncorrelated with ${\rm d}W(t')$ for $t' > t$.
The equation is solved via the ansatz
\begin{equation}
 x(t) = e^{{\cal L}t}y(t).
\end{equation}
We find 
\begin{eqnarray}
   {\rm d}x(t) & = & {\cal L} e^{{\cal L}t} y(t) \,{\rm d}t + e^{{\cal L}t} {\rm d}y(t) \\
   & = & {\cal L} x(t) \, {\rm d}t +  e^{{\cal L}t} {\rm d}y(t).
\end{eqnarray}
Consequently, Eq.~(\ref{SMEiteration}) can be rewritten as
\begin{equation}
  {\rm d}y(t) = e^{-{\cal L}t} a(t) \,{\rm d}W(t) 
\end{equation}
which after integration (Ito-Integral) leads us to
\begin{equation}
  x(t) = e^{{\cal L}(t-t_0)} y_0  + e^{{\cal L}t} \int_{t_0}^{t} e^{-{\cal L}\tau} a(\tau)\, {\rm d}W(\tau). 
\end{equation}
The first term is proportional to the stationary state $\rho_0$ (see Section \ref{sec:CQNF}) for $t_0 \rightarrow - \infty$, which is already 
included in the zero order contribution $\rho_0$ to $\rho(t)$. We therefore drop the first term and write
\begin{equation}
  x(t) = e^{{\cal L}t} \int_{-\infty}^{t} e^{-{\cal L}\tau} a(\tau) \,{\rm d}W(\tau). 
\end{equation}
Using the definition of ${\cal G}(t)$ given in the body of the text we find
\begin{equation}
  x(t) = \int_{-\infty}^{t} {\cal G}(t-\tau)a(\tau) \,{\rm d}W(\tau) \label{ItoConv}
\end{equation}
which is the Ito-version of a stochastic convolution integral between a usual function ${\cal G}(t)$ 
and a stochastic quantity
$a(t)$.
The evaluation of averages that contain Ito-integrals requires special care. Consider the following examples.
\begin{widetext}
The first and second order contributions to the density matrix $\rho(t)$ obtain the form
\begin{eqnarray}
 \rho_1(t)  &= &\int_{-\infty}^t {\cal G}(t-\tau) B_0(\rho_0)\, {\rm d}W(\tau) \\
 \rho_2(t) &= & \int_{-\infty}^{t_1} \int_{-\infty}^{\tau_1}  {\cal G}(t-\tau_1)  B_1(  {\cal G}(\tau_1-\tau_2)   B_0(\rho_0),B_0(\rho_0))\, {\rm d}W(\tau_2) \, {\rm d}W(\tau_1) 
\end{eqnarray}
where $\rho_2$ followed from an iterative application of Eq.~(\ref{ItoConv}).
Since $a(t)$ was assumed to be non-anticipating the averages $\langle x(t) \rangle$ are always zero and therefore $\langle \rho_1(t) \rangle = 0$ and $\langle \rho_2(t) \rangle = 0$.  
Let us check if this is for $\rho_1(t)$ and $\rho_2(t)$ consistent with a Langevin-treatment of the averages:
\begin{eqnarray}
\langle \rho_1(t) \rangle & = & \left\langle \int_{-\infty}^t {\cal G}(t-\tau) B_0(\rho_0) \Gamma(\tau)\, {\rm d}\tau \right\rangle
\nonumber \\
& = &  \int_{-\infty}^t {\cal G}(t-\tau) B_0(\rho_0) \langle \Gamma(\tau) \rangle \, {\rm d}\tau \nonumber \\
& = & 0.
\end{eqnarray}
With $B_1(\rho_1,\rho_0) = 2{\cal A}' \rho_1 - 2 \rho_0 {\rm Tr}(A \rho_1)$ and  ${\cal A}' x =  {\cal A} x - x {\rm Tr}({\cal A}\rho_0)$
we find
\begin{eqnarray}
\langle \rho_2(t) \rangle & = & \left\langle \int_{-\infty}^{t_1} \int_{-\infty}^{\tau_1}  {\cal G}(t-\tau_1)
\{ 2{\cal A'}    {\cal G}(\tau_1-\tau_2)   B_0(\rho_0) - 2\rho_0 {\rm Tr}(A {\cal G}(\tau_1-\tau_2)   B_0(\rho_0))\} \, {\rm d}W(\tau_2) \, {\rm d}W(\tau_1)  \right\rangle \nonumber \\
& = &  \int_{-\infty}^{t_1} \int_{-\infty}^{\tau_1}  {\cal G}(t-\tau_1)
\{ 2{\cal A'}    {\cal G}(\tau_1-\tau_2)   B_0(\rho_0) - 2\rho_0 {\rm Tr}(A {\cal G}(\tau_1-\tau_2)   B_0(\rho_0))\} \, \langle \Gamma(\tau_2) \Gamma(\tau_1)\rangle{\rm d}\tau_2 {\rm d}\tau_1  \nonumber\\
& = &  \int_{-\infty}^{t_1} \int_{-\infty}^{\tau_1}  {\cal G}(t-\tau_1)
\{ 2{\cal A'}    {\cal G}(\tau_1-\tau_2)   B_0(\rho_0) - 2\rho_0 {\rm Tr}(A {\cal G}(\tau_1-\tau_2)   B_0(\rho_0))\} \, 
\delta(\tau_2-\tau_1) {\rm d}\tau_2 {\rm d}\tau_1.
  \end{eqnarray}
In the last line we encounter the problem that the argument of the delta function is zero exactly when $\tau_2$ reaches the upper limit $\tau_1$ of the inner integral. Gardiner shows that within the Ito-formalism the upper limit does not contribute to the integral (compare Gardiner's Eq.~4.2.56 to 4.2.60 \cite{gardinerBOOK2009}). Consequently, the integral disappears and $\langle \rho_2(t) \rangle = 0$ follows as expected. We regard the Ito-way of treating the upper integral limit by defining ${\cal G}(t) =0$ for $t \le 0$.   
\end{widetext}
With this definition we can rewrite the Ito-integral: 
Eq.~(\ref{ItoConv})
\begin{equation}
  x(t) = \int_{-\infty}^{\infty} {\cal G}(t-\tau)a(\tau) \,{\rm d}W(\tau) \label{ItoConv2}.
\end{equation}
After introducing the symbol $\star$ for the
Ito convolution of a function and a stochastic quantity we can write Eq.~(\ref{ItoConv2}) as
\begin{equation}
 x(t) = {\cal G}(t) \star a(t).
\end{equation}
Multiple Ito integrals can appear e.g. for the evaluation of moments of $z(t)$ if the explicit expression for $z(t)$, Eq.~(\ref{eq:CQNF}), are used. They require the evaluation of 
expressions like $\langle \Gamma(t_n) \cdots \Gamma(t_1) \rangle$ which is given in Appendix \ref{app:MomentsAndCumulants}.
In this paper the multi-time moments of $z(t)$ are evaluated in Section \ref{sec:Multitime} without the need of Eq.~(\ref{eq:CQNF}).
\section{Fluctuation Dissipation Theorem}
\label{app:FDT}
Here we give a fully quantum mechanical derivation of the fluctuation dissipation theorem (FDT) using the same notations and conventions as used in the rest of the manuscript \cite{callenPR1951,kuboJPSJ1957}. Since the FDT establishes a relation between the power spectrum $S_q(\omega)$ and the susceptibility $\alpha(\omega)$, we first derive an expression for $\alpha(\omega)$.
The susceptibility $\alpha(\omega)$ of the system $\rho_0$ can be found by considering an excitation of the system given by the
interaction Hamiltonian $H_{\rm V}(t) =  - A  h(t)$ where $h(t)$ may be any time-dependent function, e.g. $h(t) = h_0 \cos (\omega_0 t)$. 
After introducing the superoperator ${\cal V}(t)\rho = \frac{i}{\hbar} [\rho, H_V(t)]$ we need to solve
\begin{equation}
 \dot{\rho} = {\cal L}\rho + \lambda {\cal V}(t)\rho.
\end{equation}
The linear response is obtained from the ansatz $\rho(t) = \rho_0(t) + \lambda \rho_1(t)$ which yields the equations
\begin{eqnarray}
 \dot{\rho}_0 & = & {\cal L} \rho_0 \\
 \dot{\rho}_1 - {\cal L} \rho_1 & = & {\cal V}(t) \rho_0.
\end{eqnarray}
The zero order contribution $\rho_0(t) = \rho_0$ is constant in equilibrium. The explicit solution for $\rho_1$ is given as a convolution
(see Appendix \ref{app:Conventions})
\begin{equation} 
 \rho_1(t) = \frac{i }{\hbar}{\cal G}(t) \ast (( A\rho_0  -  \rho_0 A)h(t)). 
\end{equation}
The expectation value for $z_1(t) = {\rm Tr}(A\rho_1(t))$ is after Fourier transformation given by
\begin{equation}
 z_1(\omega) = \frac{i }{\hbar}{\rm Tr}\left(A{\cal G}(\omega)[(A\rho_0  -  \rho_0 A)h(\omega)]\right).
\end{equation}
Since the complex susceptibility $\alpha(\omega)$ is implicitly defined via $z_1(\omega) = \alpha(\omega)h(\omega)$ we find
\begin{equation}
\alpha(\omega) = \frac{i }{\hbar}{\rm Tr}\left(A{\cal G}(\omega)[A\rho_0  -  \rho_0 A]\right).
\end{equation}
The expression for $\alpha(\omega)$ can in the absence of damping by rewritten in a form that allows for a comparison with 
$S^{\rm (L)}(\omega) = \frac{1}{2}\left({\rm Tr}\left[ A{\cal G}(\omega)(\rho_0 A + A \rho_0)\right] + {\rm c.c.}\right)$. 
Consider
\begin{equation}
\alpha(t) = \frac{i }{\hbar}{\rm Tr}\left(A{\cal G}(t)[A\rho_0  -  \rho_0 A]\right).
\end{equation}
\begin{widetext}
For $t \ge 0$ this can be expressed as ($\alpha(t) = 0$ for $t<0$)
\begin{eqnarray}
 \alpha(t) & = & {\cal N}\frac{i}{\hbar} {\rm Tr}(A e^{-iHt/\hbar} A e^{-H/k_{\rm B} T}e^{iHt/\hbar})
 - {\cal N}\frac{i}{\hbar} {\rm Tr}(A e^{-iHt/\hbar} e^{-H/k_{\rm B} T} A e^{iHt/\hbar}) \nonumber \\
 & = & {\cal N}\frac{i}{\hbar} \sum_{n,m} \langle m | A e^{-iHt/\hbar} |n \rangle \langle n |  A e^{-H/k_{\rm B} T}e^{iHt/\hbar} 
 | m \rangle
 - {\cal N}\frac{i}{\hbar} \sum_{n,m} \langle m | A e^{-iHt/\hbar} e^{-H/k_{\rm B} T} |n \rangle \langle n |  A e^{iHt/\hbar} 
 | m \rangle        \nonumber \\
 & = & {\cal N}\frac{i}{\hbar} \sum_{n,m} |A_{m,n}|^2 e^{-i(\omega_n - \omega_m)t}
                 \left( e^{-\hbar \omega_m/k_{\rm B} T} - e^{-\hbar \omega_n/k_{\rm B} T} \right) \nonumber \\
                  & = & {\cal N}  \frac{i}{\hbar}\sum_{m,n} |A_{m,n}|^2 e^{-i(\omega_n - \omega_m)t} 
                  e^{-\hbar \omega_n/k_{\rm B} T} (e^{\hbar(\omega_n-\omega_m)/k_{\rm B}T} -1),
\end{eqnarray}
where we expressed the thermal equilibrium via the canonical distribution $\rho_0 = {\cal N}e^{-H/k_{\rm B} T}$ with the normalization factor ${\cal N}^{-1} = {\rm Tr}(e^{-H/k_{\rm B}T})$. The frequencies $\omega_n$ are related to the eigenvalues of $H$ via $H | n\rangle
= \hbar \omega_n | n\rangle$.
We eventually find
\begin{equation}
  \alpha(\omega) =  {\cal N}  \frac{i}{\hbar}\sum_{m,n} |A_{m,n}|^2 \left(\pi \delta(\omega - (\omega_n - \omega_m)) +
  \frac{i}{\omega - (\omega_n - \omega_m)}\right) e^{-\hbar \omega_n/k_{\rm B} T}  \left(e^{\hbar(\omega_n-\omega_m)/k_{\rm B} T} - 1\right) \label{FTD_Diss}
\end{equation}
A very similar calculation for $S^{\rm (L)}(\omega)$ gives
\begin{equation}
S^{\rm (L)}(\omega) = {\cal N}\sum_{m,n} |A_{m,n}|^2 \pi \delta(\omega - (\omega_n - \omega_m)) 
  e^{-\hbar \omega_n/k_{\rm B} T}  \left(1+e^{\hbar(\omega_n-\omega_m)/k_{\rm B}T}\right). \label{FTD_Noise}
\end{equation}
The FDT
\begin{equation}
 {\rm Im} \,\alpha(\omega) = \frac{1}{\hbar}S^{\rm (L)}(\omega)  \frac{1- e^{-\hbar\omega/k_{\rm B} T}}{1+e^{-\hbar \omega/k_{\rm B} T}}
\end{equation}
follows from a comparison of Eq.~(\ref{FTD_Noise}) and  Eq.~(\ref{FTD_Diss}).

\section{Fourth order cumulant in the time domain}
\label{AppC4}
Here we show that the fourth order time domain cumulant is given by
\begin{equation}
 \tilde{C}_4(z(t_4),z(t_3),z(t_2),z(t_1)) = \beta^8\sum_{\text{prm. $t_j$}} {\rm Tr}[{\cal A}'{\cal G}'(t_4-t_3){\cal A}'{\cal G}'(t_3-t_2){\cal A}'{\cal G}'(t_2-t_1){\cal A}' \rho_0]. \label{eq:tildeC4}
\end{equation}
Starting from the general fourth order cumulant $C_4$ we show that $\tilde{C}_4 = C_4$ 
 \begin{eqnarray}
 C_4 & = & \langle x_4 x_3 x_2 x_1\rangle \nonumber \\
  & & - \langle x_4\rangle \langle x_3 x_2 x_1\rangle - \langle x_3\rangle \langle x_4 x_2 x_1\rangle -
  \langle x_2\rangle \langle x_4 x_3 x_1\rangle - \langle x_1\rangle \langle x_4 x_3 x_2\rangle \nonumber \\
 & &  -   \langle x_4 x_3\rangle \langle x_2 x_1\rangle - \langle x_4 x_2\rangle \langle x_3 x_1\rangle -
  \langle x_4 x_1\rangle \langle x_3 x_2\rangle \nonumber \\
  & & + 2 \langle x_4 x_3\rangle \langle x_2 \rangle \langle x_1\rangle + 2\langle x_4 x_2\rangle \langle x_3 \rangle \langle x_1\rangle + 2 \langle x_4 x_1\rangle \langle x_3 \rangle \langle x_2\rangle + 2 \langle x_3 x_2\rangle \langle x_4 \rangle \langle x_1\rangle \nonumber \\
   & & +  2 \langle x_3 x_1\rangle \langle x_4 \rangle \langle x_2\rangle + 2 \langle x_2 x_1\rangle \langle x_4 \rangle \langle x_3\rangle - 6 \langle x_4 \rangle \langle x_3\rangle \langle x_2 \rangle \langle x_1\rangle. \label{eq:Cumulant4} 
\end{eqnarray}     
\end{widetext}
We find for the time domain cumulant (here we use a short notation that should be self explanatory)
\begin{eqnarray}
 \beta^{-8} C_4 & = & {\rm Sum}(\alpha_1 + \alpha_2 + \alpha_3 + \alpha_4 + \alpha_5) \nonumber \\
\end{eqnarray}
where
\begin{eqnarray} 
 \alpha_1 & = & T(A G_{43} A G_{32} A G_{21} A \rho_0) \nonumber \\
 \alpha_2 & = & -T(A \rho_0) T( A G_{32} A G_{21} A \rho_0) \nonumber \\
 \alpha_3 & = & -\frac{1}{2} T(A G_{43} A \rho_0) T(A G_{21} A \rho_0) \nonumber \\
 \alpha_4 & = & T(A G_{43} A \rho_0) T(A \rho_0)^2 \nonumber \\
 \alpha_5 & = & - \frac{1}{4} T(A \rho_0)^4.
\end{eqnarray}
Similar to our calculation for $C_3$ (see Section \ref{sec:multicum}) the sum over all permutations of the time-indices guarantees that all terms of Eq. (\ref{eq:Cumulant4}) are correctly represented.
We now express $G_{ij}$ by $G_{ij}' + \theta_{ij}G_0$ where $\theta_{ij}G_0$ represents $\Theta(t_i - t_j){\cal G}_0$
and find after expansion and the use of the relation $ \theta_{ij}G_0 X = \rho_0 T(X) \theta_{ij}$ 
[compare Eq. (\ref{eq:traceproperty})] 
\begin{eqnarray}
\alpha_1 & = & T(A G'_{43} A G'_{32} A G'_{21} A \rho_0) \nonumber \\
 & & + T(A \rho_0) T(A G'_{32} A G'_{21} A \rho_0) \theta_{43} \nonumber \\
 & & + T(A G'_{43} A \rho_0) T(A G'_{21} A \rho_0) \theta_{32} \nonumber \\
 & & +  T(A G'_{43} A G'_{32} A \rho_0) T(A \rho_0) \theta_{21} \nonumber \\
 & & + T(A \rho_0)^2 T(A G'_{43}) \theta_{32} \theta_{21} \nonumber \\
  & & + T(A \rho_0)^2 T(A G'_{32}) \theta_{43} \theta_{21} \nonumber \\
   & & + T(A \rho_0)^2 T(A G'_{21}) \theta_{43} \theta_{32} \nonumber \\
   & & + T(A \rho_0)^4 \theta_{43} \theta_{32} \theta_{21},
\end{eqnarray}
\begin{eqnarray}
\alpha_2 & = & - T(A \rho_0) T(A G'_{32} A G'_{21} A \rho_0) \nonumber \\
 & & - T(A \rho_0)^2 T(A G'_{32} A   \rho_0) \theta_{21} \nonumber \\ 
  & & - T(A \rho_0)^2 T(A G'_{21} A   \rho_0) \theta_{32} \nonumber \\
  & & - T(A \rho_0)^4 \theta_{32} \theta_{21},
\end{eqnarray}   
\begin{eqnarray}
 \alpha_3 & = & - T(A G'_{43} A \rho_0)  T(A G'_{21} A \rho_0)/2 \nonumber \\
   &  & - T(A \rho_0)^2  T(A G'_{21} A \rho_0) \theta_{43}/2 \nonumber \\
    &  & - T(A \rho_0)^2  T(A G'_{43} A \rho_0) \theta_{21}/2 \nonumber \\
    & & - T(A \rho_0)^4 \theta_{43}\theta_{21}/2,
\end{eqnarray}
\begin{eqnarray}
 \alpha_4 & = & T(A G'_{43} A \rho_0) T(A \rho_0)^2 \nonumber \\
 & & + T(A \rho_0)^4 \theta_{43},
 \end{eqnarray}
and
\begin{eqnarray}
 \alpha_5 & = & - T(A \rho_0)^4 /4.
\end{eqnarray}
In the following we denote the $j$th term of $\alpha_i$ as $\alpha_{ij}$.
The equation $\beta^{-8} C_4  =  {\rm Sum}(\alpha_1 + \alpha_2 + \alpha_3 + \alpha_4 + \alpha_5)$ simplifies by noting that
${\rm Sum}(\alpha_{13} + \alpha_{31}) = 0$ when the time orders of all possible permutations of $t_4$, $t_3$, $t_2$, and $t_1$ are regarded. 
We also find that the terms proportional to $T(A\rho_0)^4$ cancel as ${\rm Sum}(\alpha_{18} + \alpha_{24} + \alpha_{34} + \alpha_{42} + \alpha_{51}) = 0$. 
After collecting terms this leaves us with
\begin{eqnarray}
\beta^{-8} C_4 & = & {\rm Sum} [T(A G'_{43} A G'_{32} A G'_{21} A \rho_0) \nonumber \\
 & & + T(A \rho_0) T(A G'_{32} A G'_{21} A \rho_0) \theta_{43} \nonumber \\
  & & + T(A \rho_0) T(A G'_{43} A G'_{32} A  \rho_0) \theta_{21} \nonumber \\
  & & - T(A \rho_0) T(A G'_{32} A G'_{21} A  \rho_0) \nonumber \\
 & & + T(A \rho_0)^2 T(A G'_{41} A \rho_0)(\theta_{12}\theta_{23} + \theta_{34}\theta_{12} + \theta_{23}\theta_{34} \nonumber \\
& &  - \theta_{12} - \theta_{34} - \theta_{23}/2 - \theta_{23}/2 + 1) ].
\end{eqnarray}
Checking all possible time orders of the last term we find that only the time order $t_4>t_3>t_2>t_1$ leads to a non-zero contribution
that can conveniently be written as ${\rm Sum}(T(A \rho_0)^2 T(A G'_{43} G'_{32} G'_{21} A \rho_0))$ where the product of propagators $G'$ is only non-zero for the required time order. We find
\begin{eqnarray}
 \beta^{-8} C_4 & = & {\rm Sum} [T(A G'_{43} A G'_{32} A G'_{21} A \rho_0) \nonumber \\
 & & + T(A \rho_0) T(A G'_{32} A G'_{21} A \rho_0) \theta_{43} \nonumber \\
  & & + T(A \rho_0) T(A G'_{43} A G'_{32} A  \rho_0) \theta_{21} \nonumber \\
  & & - T(A \rho_0) T(A G'_{32} A G'_{21} A  \rho_0) \nonumber \\
 & & + T(A \rho_0)^2 T(A G'_{43} G'_{32} G'_{21} A \rho_0). \label{C4final}
\end{eqnarray}
The expression for $\tilde{C}_4$ is now shown to obtain the form of Eq.~(\ref{C4final})
\begin{eqnarray}
 \beta^{-8} \tilde{C}_4 & = & {\rm Sum} T( A   G'_{43} (A - T(A\rho_0))  \nonumber \\
& & \times G'_{32} (A - T(A\rho_0))   G'_{21} A \rho_0) \nonumber \\
 & = &    {\rm Sum} [ T(A G'_{43} A G'_{32} A G'_{21} A \rho_0) \nonumber \\
  & & -  T(A \rho_0) T(A G'_{43}  G'_{32} A G'_{21} A \rho_0) \nonumber \\
    & & - T(A \rho_0) T(A G'_{43}  A G'_{32}  G'_{21} A \rho_0) \nonumber \\
    & & +  T(A \rho_0)^2 T(A G'_{43} G'_{32} G'_{21} A \rho_0)].
\end{eqnarray}
where we could replace the first and last ${\cal A}'$ of Eq. (\ref{eq:tildeC4}) by ${\cal A}$ with the same arguments used for the third order case [see discussion before Eq. (\ref{eq:C3}].  
We can replace $G'_{43} G'_{32}$ in the last line by $G'_{42}(1-\theta_{34}-\theta_{23})$ considering the required time order
$t_4 > t_3 > t_2$ for a non-zero contribution under the sum.
Similarly  $G'_{32}  G'_{21}$ in the last but one line can be replaced.
We find 
\begin{eqnarray}
 \beta^{-8} \tilde{C}_4  & = &    {\rm Sum} [T(A G'_{43} A G'_{32} A G'_{21} A \rho_0) \nonumber \\
  & & -   T(A \rho_0) T(A G'_{42} A G'_{21} A \rho_0) (1-\theta_{34}-\theta_{23}) \nonumber \\
    & & -   T(A \rho_0) T(A G'_{43}  A G'_{31} A \rho_0) (1-\theta_{23}-\theta_{12}) \nonumber \\
    & & +  T(A \rho_0)^2 T(A G'_{43} G'_{32} G'_{21} A \rho_0)].
\end{eqnarray}
The summation over permutations allows us to exchange the index 2 and 3 in the last but one line to arrive at 
\begin{eqnarray}
 \beta^{-8} \tilde{C}_4  & = &    {\rm Sum} [T(A G'_{43} A G'_{32} A G'_{21} A \rho_0) \nonumber \\
  & & - T(A \rho_0) T(A G'_{42} A G'_{21} A \rho_0) \nonumber \\
 & & \times (1-\theta_{34}-\theta_{23} + 1 - \theta_{32} -\theta_{13}) \nonumber \\
       & & +   T(A \rho_0)^2 T(A G'_{43} G'_{32} G'_{21} A \rho_0)].
\end{eqnarray}
where after noting that $1 - \theta_{32} - \theta_{23} = 0$ and exchanging indices (where required) we arrive at the 
same expression as for $C_4$ [Eq.~(\ref{C4final})] which eventually leaves us with $\tilde{C}_4 = C_4$.  
% \bibliographystyle{prsty} 
%\bibliography{spinnoise}

\begin{thebibliography}{10}

\bibitem{ubbelohdeNATCOMM2012}
N. Ubbelohde {\it et~al.}, Nat. Commun. {\bf 3},  612  (2012).

\bibitem{gustavssonPRB2007}
S. Gustavsson {\it et~al.}, Phys. Rev. B {\bf 75},  075314  (2007).

\bibitem{neumannBOOK1955}
J. von Neumann, {\em Mathematical Foundations of Quantum Mechanics} (Princeton
  University Press, Princeton, 1955).

\bibitem{misraJMP1977}
B. Misra and E.~C.~G. Sudarshan, J. Math. Phys. {\bf 18},  756  (1977).

\bibitem{aharonovPRL1988}
Y. Aharonov, D.~Z. Albert, and L. Vaidman, Phys. Rev. Lett. {\bf 60},  1351
  (1988).

\bibitem{korotkovPRB2001}
A.~N. Korotkov, Phys. Rev. B {\bf 63},  085312  (2001).

\bibitem{barchielliNC1982}
A. Barchielle, L. Lanz, and G.~M. Prosperi, Nuovo Cimento {\bf 72B},  79
  (1982).

\bibitem{belavkinConf1987}
V. Belavkin,  in {\em Information Complexity and Control in Quantum Physics},
  Vol.~294 of {\em International Centre for Mechanical Sciences}, edited by A.
  Blaquiere, S. Diner, and G. Lochak (Springer, Vienna, 1987), p.\ 311.

\bibitem{barchielliBOOK2009}
A. Barchielli and M. Gregoratti, {\em Quantum Trajectories and Measurements in
  Continuous Time: The Diffusive Case}, {\em Lecture Notes in Physics 782}
  (Springer, Berlin Heidelberg, 2009).

\bibitem{boutenSIAM2007}
L. Bouten, R. {van Handel}, and M.~R. James, SIAM J. Control Optim. {\bf 46},
  2199  (2007).

\bibitem{jacobsCP2006}
K. Jacobs and D.~A. Steck, Contemp. Phys. {\bf 47},  279  (2006).

\bibitem{diosiPLA1988}
L. Diosi, Phys. Lett. A {\bf 129},  419  (1988).

\bibitem{gagenPRA1993}
M.~J. Gagen, H.~M. Wiseman, and G.~J. Milburn, Phys. Rev. A {\bf 48},  132
  (1993).

\bibitem{korotkovPRB1999}
A.~N. Korotkov, Phys. Rev. B {\bf 60},  5737  (1999).

\bibitem{goanPRB2001}
H.-S. Goan, G.~J. Milburn, H.~M. Wiseman, and H.~B. Sun, Phys. Rev. B {\bf 63},
   125326  (2001).

\bibitem{oreshkovPRL2005}
O. Oreshkov and T.~A. Brun, Phys. Rev. Lett. {\bf 95},  110409  (2005).

\bibitem{liPRL2016}
F. Li and N.~A. Sinitsyn, Phys. Rev. Lett. {\bf 116},  026601  (2016).

\bibitem{cavesPRD1986}
C.~M. Caves, Phys. Rev. D {\bf 33},  1643  (1986).

\bibitem{menskyPLA1994}
M. Mensky, Phys. Lett. A {\bf 196},  159  (1994).

\bibitem{bednorzNJP2012}
A. Bednorz, W. Belzig, and A. Nitzan, New J. Physics {\bf 14},  013009  (2012).

\bibitem{glazovPRB2012}
M.~M. Glazov and E.~L. Ivchenko, Phys. Rev. B {\bf 86},  115308  (2012).

\bibitem{landauBOOKstat}
L. Landau and E. Lifshitz, {\em Statistical Physics} (Elsevier Science,
  Amsterdam, 2013), No.~Bd. 5.

\bibitem{kuboJPSJ1957}
R. Kubo, J. Phys. Soc. Jpn. {\bf 12},  570  (1957).

\bibitem{gardinerBOOK2009}
C. Gardiner, {\em Stochastic Methods}, fourth ed. (Springer, Berlin Heidelberg,
  2009).

\bibitem{aleksandrovJETP1981}
E.~B. Aleksandrov and V.~S. Zapasskii, Sov. Phys. JETP {\bf 54},  64  (1981).

\bibitem{crookerNATURE2004}
S.~A. Crooker, D.~G. Rickel, A.~V. Balatsky, and D.~L. Smith, Nature {\bf 431},
   49  (2004).

\bibitem{oestreichPRL2005}
M. Oestreich, M. R\"omer, R.~J. Haug, and D. H\"agele, Phys. Rev. Lett. {\bf
  95},  216603  (2005).

\bibitem{mullerPHYSICAE2010}
G.~M. M\"uller, M. Oestreich, M. R\"omer, and J. H\"ubner, Physica E {\bf 43},
  569  (2010).

\bibitem{aleksandrovJP2011}
E.~B. Aleksandrov and V.~S. Zapasskii, J. Phys. Conf. Ser. {\bf 324},  012002
  (2011).

\bibitem{hubnerPSSB2014}
J. H\"ubner, F. Berski, R. Dahbashi, and M. Oestreich, Phys. Status Solidi B
  {\bf 251},  1824  (2014).

\bibitem{glazovJETP2015}
M.~M. Glazov, J. Exp. Theor. Phys. {\bf 122},  472  (2015).

\bibitem{sinitsynRPP2016}
N.~A. Sinitsyn and Y.~V. Pershin, Rep. Prog. Phys. {\bf 79},  106501  (2016).

\bibitem{bussPRB2016}
J.~H. Bu{\ss} {\it et~al.}, Phys. Rev. B {\bf 93},  155204  (2016).

\bibitem{braunPRB2007}
M. Braun and J. K\"onig, Phys. Rev. B {\bf 75},  085310  (2007).

\bibitem{clerkRMP2010}
A.~A. Clerk {\it et~al.}, Rev. Mod. Phys. {\bf 82},  1155  (2010).

\bibitem{smirnovPRB2014}
D.~S. Smirnov and M.~M. Glazov, Phys. Rev. B {\bf 90},  085303  (2014).

\bibitem{smirnovPRB2017}
D.~S. Smirnov, B. Reznychenko, A. Auffeves, and L. Lanco, Phys. Rev. B {\bf
  96},  165308  (2017).

\bibitem{liuNJP2010}
R.~B. Liu {\it et~al.}, New J. Physics {\bf 12},  013018  (2010).

\bibitem{zapasskiiPRL2013}
V.~S. Zapasskii {\it et~al.}, Phys. Rev. Lett. {\bf 110},  176601  (2013).

\bibitem{liNJP2013}
F. Li, A. Saxena, D. Smith, and N.~A. Sinitsyn, New J. Physics {\bf 15},
  113038  (2013).

\bibitem{liPRA2016}
F. Li, S.~A. Crooker, and N.~A. Sinitsyn, Phys. Rev. A {\bf 93},  033814
  (2016).

\bibitem{starosielecRSI2010}
S. Starosielec, R. Fainblat, J. Rudolph, and D. H\"agele, Rev. Scientific
  Instrum. {\bf 81},  125101  (2010).

\bibitem{nikiasIEEE1993}
C.~L. Nikias and J.~M. Mendel, IEEE Signal Proc. Mag. {\bf 10},  10  (1993).

\bibitem{brillingerAMS1965}
D.~R. Brillinger, Ann. Math. Statist. {\bf 36},  1351  (1965).

\bibitem{efimovQJRMS2001}
V.~V. Efimov and M.~V. Shokurov, Q. J. R. Meteorol. Soc. {\bf 127},  1707
  (2001).

\bibitem{balkPRX2018}
A.~L. Balk {\it et~al.}, Phys. Rev. X {\bf 8},  031078  (2018).

\bibitem{annabestaniJMR2015}
R. Annabestani, D.~G. Cory, and J. Emerson, J. Magn. Reson. {\bf 252},  94
  (2015).

\bibitem{gillespieAJP1995}
D.~T. Gillespie, Am.\ J.\ Phys. {\bf 64},  225  (1996).

\bibitem{barchielliQM2013}
A. Barchielli and M. Gregoratti, Quantum Meas. Quantum Metrol. {\bf 1},  34
  (2013).

\bibitem{lindbladSPRINGER1976}
G. Lindblad, Commun. Math. Phys. {\bf 48},  119  (1976).

\bibitem{laxPR1963}
M. Lax, Phys. Rev. {\bf 129},  2342  (1963).

\bibitem{carmichaelBOOK1993}
H. Carmichael, {\em An Open Systems Approach to Quantum Optics}, {\em Lecture
  Notes in Physics} (Springer, Berlin Heidelberg, 1993).

\bibitem{mullerPRL2008}
G.~M. M\"uller {\it et~al.}, Phys. Rev. Lett. {\bf 101},  206601  (2008).

\bibitem{poltavtsevPRB2014}
S.~V. Poltavtsev {\it et~al.}, Phys. Rev. B {\bf 89},  081304  (2014).

\bibitem{yangRPP2017}
W. Yang, W.-L. Ma, and R.-B. Liu, Rep. Prog. Phys. {\bf 80},  016001  (2017).

\bibitem{langevinCR1908}
P. Langevin, C. R. Acad. Sci. (Paris) {\bf 146},  530  (1908).

\bibitem{uhlenbeckPR1930}
G.~E. Uhlenbeck and L.~S. Ornstein, Phys. Rev. {\bf 36},  823  (1930).

\bibitem{blockPRB82}
D. Block, A. Herve, and R.~T. Cox, Phys. Rev. B {\bf 25},  6049  (1982).

\bibitem{kitagawaPRA1993}
M. Kitagawa and M. Ueda, Phys. Rev. A {\bf 47},  5138  (1993).

\bibitem{fisherPLMS1930}
R.~A. Fisher, Proc. Lond. Math. Soc. 2nd Ser. {\bf 30},  199  (1930).

\bibitem{atalayaPRA2018}
J. Atalaya {\it et~al.}, Phys. Rev. A {\bf 97},  020104(R)  (2018).

\bibitem{tilloyPRA2018}
A. Tilloy, Phys. Rev. A {\bf 98},  010104(R)  (2018).

\bibitem{marcosNJP2010}
D. Marcos, C. Emary, T. Brandes, and R. Ajuado, New J. Physics {\bf 12},
  123009  (2010).

\bibitem{hubenerPRL2013}
R. H\"ubener, A. Mari, and J. Eisert, Phys. Rev. Lett. {\bf 110},  040401
  (2013).

\bibitem{forguesSR2013}
J.-C. Forgues {\it et~al.}, Sci. Rep. {\bf 3},  2869  (2013).

\bibitem{callenPR1951}
H.~B. Callen and T.~A. Welton, Phys. Rev. {\bf 83},  34  (1951).

\end{thebibliography}

\end{document}